  \documentclass[manuscript]{emulateapj}

\usepackage{apjfonts}

\slugcomment{to appear in the Astrophysical Journal}

\shorttitle{Light Curve of GQ Muscae}
\shortauthors{Hachisu et al.}

\begin{document}

\title{A Universal Decline Law of Classical Novae. III. GQ Mus 1983}


\author{Izumi Hachisu}
\affil{Department of Earth Science and Astronomy,
College of Arts and Sciences, University of Tokyo,
Komaba, Meguro-ku, Tokyo 153-8902, Japan}
\email{hachisu@ea.c.u-tokyo.ac.jp}


\author{Mariko Kato}
\affil{Department of Astronomy, Keio University,
Hiyoshi, Kouhoku-ku, Yokohama 223-8521, Japan}
\email{mariko@educ.cc.keio.ac.jp}

\and

\author{Angelo Cassatella}
\affil{INAF, Istituto di Fisica dello Spazio Interplanetario,
Via del Fosso del Cavaliere 100, 00133 Rome, Italy}
\email{cassatella@fis.uniroma3.it}




\begin{abstract}
We present a unified model of infrared (IR), optical, ultraviolet (UV), and
X-ray light curves for the 1983 outburst of GQ Muscae (Nova Muscae 1983)
and estimate its white dwarf (WD) mass.  Based on an optically thick
wind model of nova outbursts, we model the optical and IR light curves
with free-free emission, and the UV 1455 \AA\ and supersoft X-ray light
curves with blackbody emission.  The best fit model that reproduces
simultaneously the IR, optical, UV 1455 \AA, and supersoft X-ray
observations is a $0.7 \pm 0.05~M_\sun$ WD for an assumed chemical
composition of the envelope, $X=0.35-0.55$, $X_{\rm CNO} =0.2-0.35$,
and $Z = 0.02$, by mass weight.   The mass lost by the wind is 
estimated to be $\Delta M_{\rm wind} \sim 2 \times 10^{-5} M_\odot$.
We provide a new determination of the reddening, $E(B-V) = 0.55 \pm 0.05$,
and of the distance, $\sim 5$~kpc.  Finally, we discuss the
strong UV flash that took place on JD 2,445,499 (151 days after the
outburst).
\end{abstract}




\section{Introduction}
Despite their overall similarity, the optical light curves of classical
novae show a wide variety of timescales and shapes
\citep[e.g.,][]{pay57}. Various empirical time-scaling laws have been
proposed in the attempt to recognize common patterns and to unify
the nova light curves.  For example, \citet{mcl42} proposed
a compression of the time scale obtained by normalizing the time to
$t_m$, the time for the brightness to decrease by $m$ magnitudes
(usually $m=2$ or 3).  The results were however unsatisfactory
because slow novae reached their late stages relatively earlier
than faster novae.  A different approach was adopted by \citet{vor48}
who, based on a large collection of nova light curves,
found that $m(t)$, the visual magnitude at time $t$, is
conveniently represented, on average, by $m(t)= m_0 + b_1
\log (t-t_0)$ at early stages and $m(t)= m_1 + b_2 \log (t-t_0)$
at later stages with $b_1 = 2.5$ and $b_2 > b_1$.

The underlying problem of both the above approaches is that, during the
decline phase from maximum, the flux in the visual band, initially dominated
by free--free emission, becomes more and more affected by the increasing
contribution from the emission lines, which finally dominates.  The presence
of these two competing and heterogeneous emission mechanisms is most likely
the main cause for the difficulty to find a suitable normalization parameter
for the nova light curves.  A more appropriate way to monitor the evolution of
the visual continuum is to use the Str\"omgren $y$ filter, which is
designed to avoid strong emission lines and, in particular, the [\ion{O}{3}]
line \citep[e.g.,][]{kal86, loc76}.  In the case of V1668 Cyg, \citet{kal86}
has indeed shown that the continuum $y$ flux declines much faster than the
H$\beta+$[\ion{O}{3}] fluxes.  Consequently, the decline rate is much faster
in the $y$ band than in the visual band. This was demonstrated by
\citet{kal86}, who showed that the $y$ magnitude of V1668 Cyg declines with a
slope of $b_1= 3.5$ at early stages and it then changes to $b_2= 6.5$ at later
stages, values that are sensibly larger than those reported by \citet{vor48}.

The observations of V1500 Cyg by \citet{gal76} suggest that the visual and
infrared continua are well represented by free-free emission during the early
decay phase.  The same applies to GQ Mus: \citet{kra84} reported that, during
the early 40 days (diffuse enhanced and Orion-phases), the energy spectrum of
$0.3 - 10\mu$ was well represented by optically thin free-free emission
\citep[see also][]{din86}.  The flux in the optical and infrared
appears to decay with time as $F_\lambda \propto t^{-\alpha}$
in various speed classes of novae 
\citep[e.g.,][for V1500 Cyg]{enn77, woo97}.

An interesting attempt to unify the light curves of novae was proposed
by \citet{ros99a, ros99b}, who suggested that the main parameter
determining the shape of the light curve is the radius of the ejected shell,
$\log R_{\rm shell} = \log(t) + \log (v)$, where $t$ is the time from
optical maximum, and $v$ is the velocity of the ejecta.  He plotted various
nova light curves in the $m-\log R_{\rm shell}$ diagram and found that light
curves overlapped each other.  The rather large scatter of the data in
that diagram is probably due to the fact that the expansion
velocity of novae varies with time \citep[see][]{cas05}.

\citet{cas02} studied the time evolution of the UV continuum flux
in twelve CO and ONeMg novae and found a strong correlation between
the $t_3$ time and the time of maximum flux in the 1455 \AA\ continuum
(see Fig. 4 in their paper).  In a subsequent paper, \citet{cas05}
studied the time evolution of the UV emission lines in seven CO novae
and found a strong correlation between the line ionization potential
and the time of maximum emission normalized to the $t_3$ (Fig. 5
in their paper).  These results strongly suggest that novae do
indeed evolve following a common pattern being time-normalized
by the $t_3$ time, which mainly depends on the white dwarf mass
\citep[e.g.,][]{liv92}.

Recently, \citet[][hereafter referred as Paper I]{hac06kb} found that
the visual and IR light curves of several novae follow a universal
decline law, and interpreted that in terms of free-free emission
\citep[Paper I;][hereafter referred as Paper II]{hac07k};
in particular, the time-normalized light curves were found to be
independent of the white dwarf mass, the chemical composition
of ejecta, and wavelength.  They also showed that the UV 1455 \AA\ 
light curve, interpreted as blackbody continuum,
can also be time-normalized by the same factor as in the optical
and IR.  The authors, in the above quoted papers, determined the
white dwarf mass and other parameters for a number of relatively
well-observed classical novae.

In the present paper, we apply the above universal decline law to
the classical nova GQ Muscae 1983.
GQ Mus is the first classical nova in which the
supersoft X-ray turnoff was detected with the X-ray satellite
{\it ROSAT} \citep{oge93, bal98, ori01}.  This object was also well
observed by the UV satellite {\it IUE} \citep[e.g.,][]{kra84}
and in near-IR bands of {\it JHKL} \citep[e.g.,][]{whi84}.

The next section summarizes the basic observational characteristics
of GQ Mus.  In \S 3, we revisit the UV light curves of
GQ Mus obtained with the {\it IUE} satellite.  In \S 4, we briefly
introduce our method based on our optically thick wind model.
Light curve fittings of GQ Mus in the X-ray, UV, optical, and
near IR bands are shown in \S 5.  Discussion and conclusions follow
in \S\S 6 and 7, respectively.  Finally, in the Appendix,
we assess the problem of the contribution of the emission
lines to the visual and infrared photometric bands for GQ Mus
and for other well known novae: V1500 Cyg, V1668 Cyg, and V1974 Cyg.

\section{Basic observational summary of GQ Muscae 1983}
\label{light-curve}
GQ Mus was discovered by Liller on 1983 January 18.14 UT
(JD 2,445,352.64) at $m_V \approx 7.2$ \citep{lil83}.
\citet{bat83} reported two pre-discovery magnitudes, i.e.,
$m_V \approx 7.6$ obtained by Gainsford on 1983 January 15.597 UT
(2,445,350.097) and $m_V \approx 7.9$ obtained by Pattie
on 1983 January 15.628 UT (2,445,350.128).
\citet{kra84} suggested that the outburst took place 3--4 days
before the discovery.  In absence of precise estimates,
we assume that the outburst took place at $t_{\rm OB} =$
~JD~ 2,445,348.0 (1983 January 13.5 UT), i.e., 4.6 days
before the discovery by Liller on January 18.14, and  adopt
$t_{\rm OB}=$ JD 2,445,348.0 as day zero in the following analysis.
 
Early optical {\it UBVRI} and near infrared {\it JHKL} observations
of GQ Mus show a fast evolution with $t_2 \sim 18$ days and
$t_3 = 48$ days \citep{whi84}.  \citet{kra84} observed GQ Mus over
a wide wavelength range from 0.12 -- 10 $\mu$ during the early 40 days
(diffuse enhanced and Orion-phases).  These authors reported
an extraordinary large amplitude brightening by $\Delta m_V 
\sim 14$ mag, with respect to the prenova magnitude of $m_V
\gtrsim 21$.  After about 4 months of slow decline, the brightness
stabilized in the range $m_V \sim 10-11$ for about 400 days.
After that, the nova gradually declined to $m_V \sim 14$
for another 400 days (see figures below).

In this paper we will use the optical and IR photometric data from
\citet{whi84} and \citet{kra84}, which cover about 300 days after discovery.
No other systematic photometric observations are available at later stages
except for the visual magnitude measurements ($V_{\rm FES}$) obtained with
the Fine Error Sensor (FES) monitor on board {\it IUE}, which cover 500 days
after discovery, and the visual photometric data collected by RASNZ (Royal
Astronomical Society of New Zealand) and by AAVSO (American Association of
Variable Star Observers), which cover 1000 days after discovery.

\citet{dia89} determined a photometric orbital period of
0.0594 days for GQ Mus, the shortest known for a classical nova.
Later on, \citet{dia94} revised the ephemeris of GQ Mus based
on the 1989 and 1990 photometric observations, i.e.,
\begin{equation}
{\rm HJD}_{\rm min} = 2,447,843.4721 + 0.0593650 \times E .
\end{equation}
We have adopted this ephemeris for our binary model.
\citet{dia89, dia94} also suggested that GQ Mus is a polar system
because of the presence of X-ray emission and high
(\ion{He}{2} $\lambda 4686$)/H$\beta$ ratio.

Based on the optical observation from 1984 March to 1988 March
(400 -- 1800 days after the outburst), \citet{kra89} suggested
that the observed progressive increase in the ionization level
of the nova shell was due to a very hot radiation source with
a temperature of $ > 4 \times 10^5$~K, increasing with time
while the bolometric luminosity remained constant.
\citet{oge87} studied the soft X-ray {\it EXOSAT} light curve
from optical maximum to about 900 days after it, and suggested
a model of a very hot white dwarf remnant with a maximum
temperature of $(2-4) \times 10^5$~K evolving at constant
bolometric luminosity.

{\it ROSAT} observations clearly show that the supersoft X-ray flux decayed
about 10 yr after the outburst \citep{oge93, sha95, ori01, bal01}.  This
information is very important because, compared with our model predictions, it
allows one to determine the white dwarf mass quite accurately (see, e.g.,
Paper I).


\begin{figure}
\epsscale{1.0}
\plotone{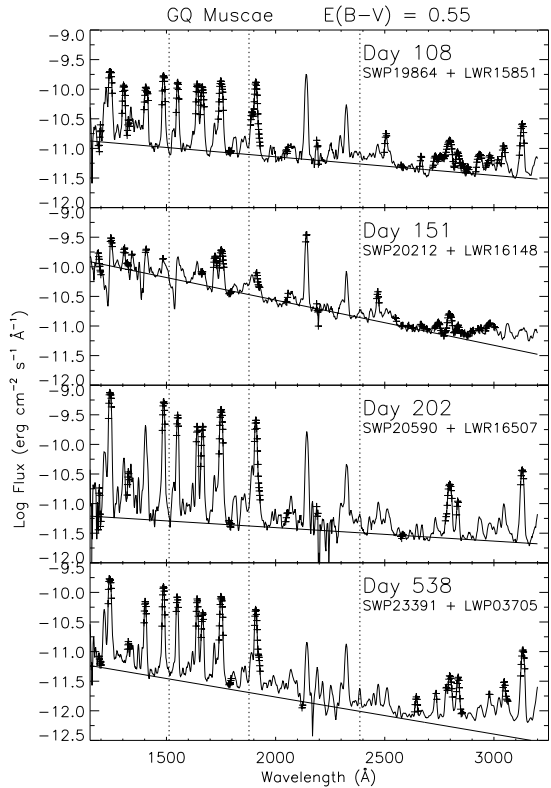}
\caption{ {\it IUE} spectra of GQ Mus obtained at four epochs,
i.e., 108, 151, 202, and 538 days after the outburst.
The spectra have been corrected for reddening using 
$E(B-V) =0.55$.  The vertical dotted lines represent the
wavelengths $\lambda\lambda$ 1512, 1878 and 2386 \AA\ 
at which the extinction law takes the same value. 
With the adopted value of reddening, the stellar continuum
underlying the many emission lines is well represented
by a straight line all over the full spectral range
except for the spectrum of day 538 above 2700 \AA,
due to the increased contribution from hydrogen Balmer
continuum.  Saturated points in the emission lines are
denoted by pluses.  Note the comparably harder spectrum
at the time of the secondary outburst on day 151.}
\label{ebvplot}
\end{figure}

\section{UV observations}
\label{uv_observations}
One of the most puzzling features of GQ Mus is its long lasting brightness in
the UV range, which made possible its regular monitoring by {\it IUE} over as
much as 11.5 yr.  Because of satellite constraints, the {\it IUE} observations
could not start before JD 2,445,385.07, i.e., 32.4 days after discovery.  The
early {\it IUE} observations of GQ Mus are discussed in detail by
\citet{kra84}.  In the following we revisit the problem of the color excess
$E(B-V)$ of GQ Mus and describe the long term evolution of the ultraviolet
continuum and of the emission lines, as well as the secondary outburst
which took place about 151 days after the main one.  The ultraviolet spectra
were retrieved from the {\it IUE} archive through the INES ({\it IUE} Newly
Extracted Spectra) system\footnote{http://ines.laeff.esa.es}, which also
provides details of the observations.  The use of {\it IUE} INES data is
particularly important for the determination of the reddening correction
because of the implementation of upgraded spectral extraction and flux
calibration procedures compared to previously published UV spectra.


\begin{figure}
\plotone{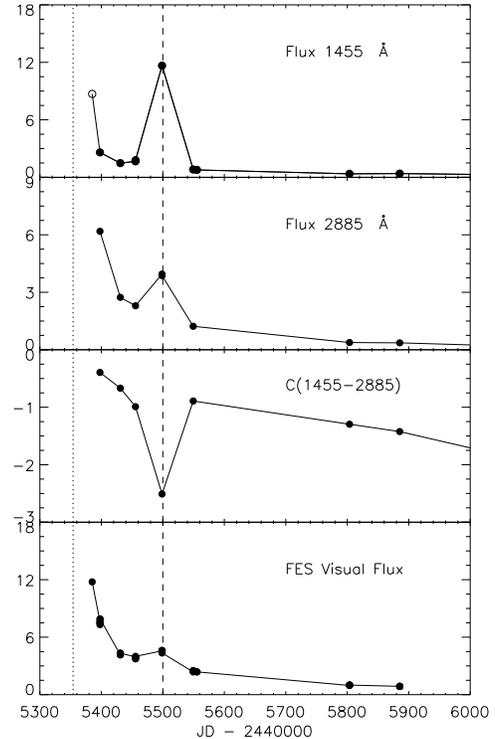}
\caption{Time evolution of the continuum fluxes at 1455 \AA\ 
and 2885 \AA, of the ultraviolet color index $C(1455 - 2885)$,
and of the visual flux.  Fluxes are in units of 
$10^{-13}$~erg~cm$^{-2}$ s$^{-1}$~\AA\ $^{-1}$, not
corrected for reddening.  Only the color index has been
corrected for reddening using $E(B-V) = 0.55$.
The vertical dotted and dashed lines indicate the date of
discovery and that of the secondary outburst, respectively.}
\label{plotcont}
\end{figure}

\subsection{The reddening correction}
The color excess of GQ Mus has been determined from the hydrogen Balmer lines
by \citet{pac85} and \citet{peq93}, who found $E(B-V) = 0.43$ and $0.50 \pm
0.05$, respectively.  About the same range of values has been reported by
\citet{kra84} and \citet{has90}, who found $E(B-V) = 0.45$ and 0.50,
respectively, based on the strength of the dust absorption band around 2175
\AA\ in the early {\it IUE} spectra.  Given the criticality of the
reddening correction, we have attempted to improve its
accuracy by using two independent methods based on the shape of the UV
continuum and on the emission line intensities.

The Galactic extinction curve \citep{sea79} shows a pronounced broad maximum
around 2175 \AA\  due to dust absorption, but it takes the same value
$X(\lambda) = A(\lambda)/E(B-V) \approx 8$ at $\lambda = 1512$, 1878,
and 2386 \AA, so that the slope of the straight line passing through the
continuum points at these wavelengths is insensitive to $E(B-V)$ in a
$(\lambda , \log F(\lambda))$ plot.  This circumstance can be used to get a
reliable estimate of $E(B-V)$ as that in which the stellar continuum becomes
closely linear in the 1512--2386 \AA\ region, and passes through the continuum
points at the above wavelengths.  From 8 pairs of short and long wavelength
{\it IUE} spectra taken during the nebular phase we have in this way found
$E(B-V) = 0.58 \pm 0.04$.  

     A different way to estimate the $E(B-V)$ color excess
consists in using the observed relative intensities
of the \ion{He}{2} 1640 \AA\ Balmer line,
and the 2734 and 3203 \AA\ Paschen recombination lines, and
compare these with theoretical ratios \citep{hum87}.
The intensity ratios $I(1640)/I(2734)$ and $I(1640)/I(3203)$
were measured in 8 pairs of short and long wavelength
{\it IUE} low resolution
spectra obtained during the nebular phase. From a comparison with the
theoretical values for an electron temperature and density of ($T_e$,
$N_e$)=(20,000 K, 10$^6$ cm$^{-3}$) we have obtained 22 independent
determinations of the color excess leading to
a mean value of $E(B-V) = 0.51 \pm 0.06$.
In the following we will adopt the error--weighted mean of the
two above determinations, i.e. $E(B-V) = 0.55 \pm 0.05$.
Examples of {\it IUE} spectra of GQ Mus corrected
with $E(B-V) = 0.55$ are reported in Figure \ref{ebvplot} for the following
days (after the outburst): 108, 151 (secondary outburst; see later), 202 and
538.


\begin{figure}
\plotone{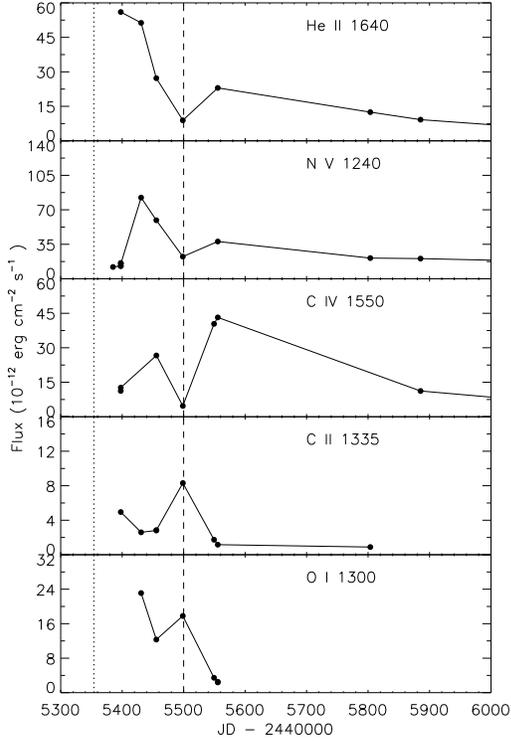}
\caption{Time evolution of the observed fluxes in the most
prominent permitted emission lines of GQ Mus, in units of
$10^{-12}$~erg~cm$^{-2}$~s$^{-1}$.
The vertical dotted and dashed lines indicate the date of
discovery and that of the secondary outburst, respectively.}
\label{plotlines}
\end{figure}


\begin{figure}
\plotone{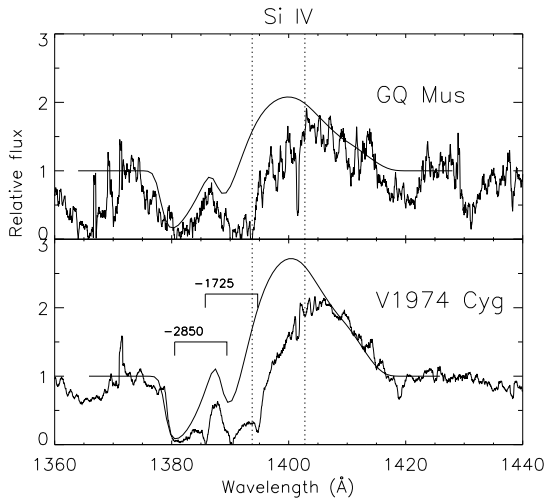}
\caption{The figure shows the P Cygni profile of the Si IV
doublet $\lambda\lambda$ 1393.74, 1402.77 \AA\ of GQ Mus
at the time of the secondary outburst on day 150.64 (upper panel)
and, for comparison, that of the Neon nova V1974 Cyg obtained
on day 45 after discovery (lower panel): the IUE images used
are SWP 20211 and SWP44390, respectively.  The observed
profiles have been fitted with theoretical P Cygni
profiles obtained through the SEI method \citep{lam87, gro89}.
Fluxes are normalized to the local continuum.  The wind terminal
velocity derived from the observed profiles is 3200 km~s$^{-1}$
in both cases.  The laboratory wavelengths of the doublet
are indicated as vertical dotted lines.  The figure indicates
also the position of the violet shifted absorption components
from the principal system at 1725 and 2850 km~s$^{-1}$ in
V 1974 Cygni \citep{cas04}.  Such components cannot be clearly
distinguished in GQ Mus due to the poor signal quality (the
spectrum is the result of a 11--point running smoothing
average).}
\label{pcygni}
\end{figure}

\subsection{Evolution of the UV continuum}
\label{uv_evolution_cont}
We have measured the mean flux in two narrow bands 20 \AA\ wide centered at
1455 \AA\ and 2855 \AA, selected to best represent the UV continuum because
little affected by emission lines (Cassatella et al. 2002). Figure
\ref{plotcont} shows the time evolution of the $F(1455$~\AA) and $F(2885$ \AA)
fluxes and of the UV color index $C(1455-2885) = -2.5
\log[F(1455$~\AA)$/F(2885$~\AA)].  The measurements were made on well exposed
low resolution large aperture spectra. Unfortunately, the {\it IUE} spectrum
of day 37 (SWP19299), very important because it was the first obtained, was
heavily saturated around 1455 \AA. To make a crude estimate of the
corresponding flux we have determined the scaling factor between this and the
next spectrum obtained on day 49 (SWP 19383), in a region where both were well
exposed (1520-1620\AA),  and assumed that the spectral slope remained the same
in the two spectra. The value so obtained, reported as an open circle in
Figure \ref{plotcont}, likely represents an upper limit to the true value
because, as shown in the same figure, the UV color was comparatively cooler in
the first observations.
Figure \ref{plotcont} reports also, for comparison,
the visual light curve obtained from the Fine Error Sensor (FES) counts on
board {\it IUE}, once corrected for FES time dependent sensitivity degradation
(see Cassatella et al. 2004 for details on the FES calibration).

It appears from Figure \ref{plotcont} that the gradual fading of the UV and
visual fluxes was interrupted, around JD 2,445,499 (day 151), by a secondary
outburst, as also noticed by \citet{has90}.
The secondary outburst had actually
the appearance of a ``UV flash'' because of its especially large amplitude
at short wavelengths. Indeed, compared with the {\it IUE} low resolution
observations obtained just before and after this event (days 108 and 202),
the UV flux increased by a factor of 9 at 1455 \AA\ and by a factor 2.2 at
2885 \AA, while the visual flux increased only by a factor of 1.5.  The
consequent hardening of the UV continuum is also reflected by the upturn of
the $C(1455-2885)$ color index, as shown in the same figure.


\begin{deluxetable*}{llllll}
\tabletypesize{\scriptsize}
\tablecaption{Chemical Abundance by Weight
\label{gqmus_chemical_abundance}}
\tablewidth{0pt}
\tablehead{
\colhead{object} &
\colhead{H} &
\colhead{CNO} &
\colhead{Ne} &
\colhead{Na-Fe} &
\colhead{reference}
}
\startdata
Solar composition &  0.7068 & 0.0140 & 0.0018 & 0.0034 & \citet{gre89} \\
GQ Mus 1983 & 0.37 & 0.24 & 0.0023  & 0.0039 & \citet{mor96a} \\
GQ Mus 1983 & 0.27 & 0.40 & 0.0034  & 0.023 & \citet{has90} \\
GQ Mus 1983 & 0.43 & 0.19 & \nodata  & \nodata & \citet{and90}
\enddata
\end{deluxetable*}


\begin{figure}
\epsscale{1.15}
\plotone{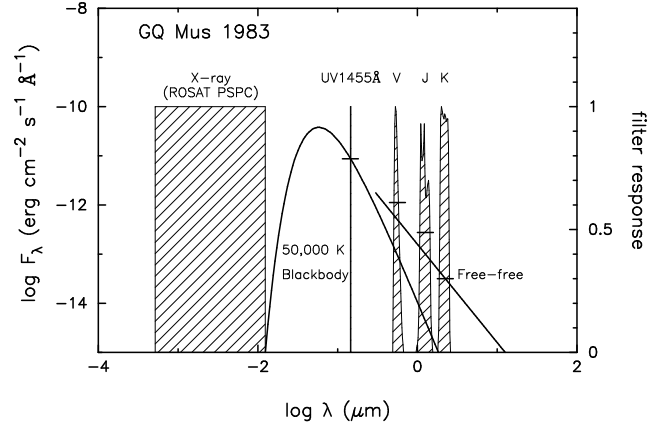}
\caption{
Schematic representation of the energy spectrum of GQ Mus
according to our model. The passbands of the photometric filters
used in this work is indicated.  The UV 1455\AA~ and supersoft
X-ray ($0.1-2.4$~keV) fluxes are calculated from a blackbody
spectrum, while the $V$, $J$, $H$, and $K$ magnitudes are from
a free-free emission spectrum as shown in the figure.  Note that
the flux scale of the dereddened blackbody and of the free--free
emission corresponds to our $0.7 ~M_\sun$ white dwarf model on
day 180 as shown in Fig.\ref{all_mass_gq_mus_x45z02c15o20_compare_uv_x},
assuming a reddening of $E(B-V)=0.55$.  Horizontal short lines
denote the energy flux in each filter on day 180; some of them
are interpolated from other observational points.
The soft X-ray flux is negligibly small on that day.
The $V$ and $J$ magnitudes are larger than those of free-free
because these bands are heavily contaminated by strong emission
lines as explained in the Appendix.}
\label{sed_uv_opt_ir_filter}
\end{figure}

\subsection{Evolution of the UV emission lines}
Also the UV emission lines suffered from important changes in coincidence
with the secondary outburst of day 151, as shown in Figure  \ref{plotlines},
which reports, as a function of time, the flux in the most prominent
permitted UV emission lines, as measured by us from the available {\it IUE}
low resolution spectra.  The figure shows that, on day 151, the flux of the
high ionization emission lines (\ion{He}{2} 1640 \AA, \ion{N}{5} 1240 \AA\
and \ion{C}{4} 1550 \AA) had decreased substantially while, on the contrary,
it increased in the low ionization resonance lines of \ion{C}{2} and
\ion{O}{1}, so  implying a drastic change of the ionization structure of
the emitting regions.

The other important feature of the secondary outburst was the appearance of a
strong P Cygni profile in the \ion{Si}{4} doublet $\lambda\lambda$ 1393.74,
1402.77 \AA, shown in Figure \ref{pcygni}. The figure reports also, for
comparison, the \ion{Si}{4} profile of the Neon Nova V1974 Cyg obtained 45
days after the outburst.  A comparison of observed with theoretical P Cygni
profiles computed with the SEI method \citep{lam87, gro89} indicate that the
terminal velocity of the wind was about the same for the two novae ($\approx$
3200 km~s$^{-1}$). The special interest of the observations in Figure
\ref{pcygni} is that GQ Mus had entered the nebular phase well before day 151,
so that the presence of P Cygni profiles was not expected to be
detectable at this stage.  Indeed, P Cygni profiles were not detectable in
an earlier {\it IUE} high resolution spectrum of day 49, despite the factor of
two longer exposure time, nor they were at later stages. This strongly
suggests that a short duration mass ejection episode took place around day
151.  This episode caused efficient recombination to take place in the
emitting region, observed as a substantial, although transitory, fainting of
the high ionization emission lines (cf.  Fig.\ref{plotlines}).  The fast
recovery of the emission line spectrum by day 202 suggests that the ejected
mass must have been small.  Also the presence of a flare at UV
wavelengths rather than in the optical (cf.  Fig.  \ref{plotcont})
requires a small opacity in the UV and then a small ejected mass.

\section{Modeling of  Nova Outbursts}
\label{model_nova_outburst}
     We present a unified model for the IR, optical, UV,
and supersoft X-ray light curves of the 1983 outburst of
GQ Mus.  As in Paper I,  our models
are based on the optically thick wind theory of nova outbursts
described in \citet{kat94h}.

\subsection{Optically thick wind model}
After a thermonuclear runaway sets in on a mass-accreting WD,
its envelope expands greatly to $R_{\rm ph} \gtrsim 100 ~R_\sun$,
where $R_{\rm ph}$ is the photospheric radius, and it then
settles onto steady-state regime.  The decay phase of the nova
can be followed by a sequence of steady state solutions 
\citep[e.g.,][]{kat94h}.  Using the same method and numerical
techniques as in \citet{kat94h}, we have followed the nova
evolution by connecting steady state solutions along the envelope
mass-decreasing sequence.

The equations of motion, radiative diffusion, and conservation
of energy are solved from the bottom of the hydrogen-rich envelope
through the photosphere under the condition that the solution
goes through a critical point of a steady-state wind.  The winds
are accelerated deep inside the photosphere so that they are
called ``optically thick winds.''  We have used updated OPAL
opacities \citep{igl96}.   One of the boundary conditions for our
numerical code consists in assuming that photons are emitted
at the photosphere as a blackbody with a photospheric temperature
$T_{\rm ph}$.  X-ray and UV fluxes are estimated directly from
the blackbody emission, but infrared and optical fluxes are
calculated from free-free emission by using physical values of
our wind solutions as mentioned below in \S\ref{free-free_light_curve}.
Physical properties of these wind solutions have already been
published \citep[e.g.,][]{hac01ka, hac01kb, hac04k, hkn96, hkn99,
hknu99, hkkm00, hac03a, kat83, kat97, kat99}.

Optically thick winds stop after a large part of the envelope is
blown in the winds.  The envelope settles into hydrostatic
equilibrium, while its mass decreases in time by nuclear burning.
Then we solve the equation of static balance instead of the
equation of motion.  When the nuclear burning decays, the WD
enters a cooling phase, in which the luminosity
is supplied with heat flow from the ash of hydrogen burning.


\begin{figure*}
\epsscale{0.85}
\plotone{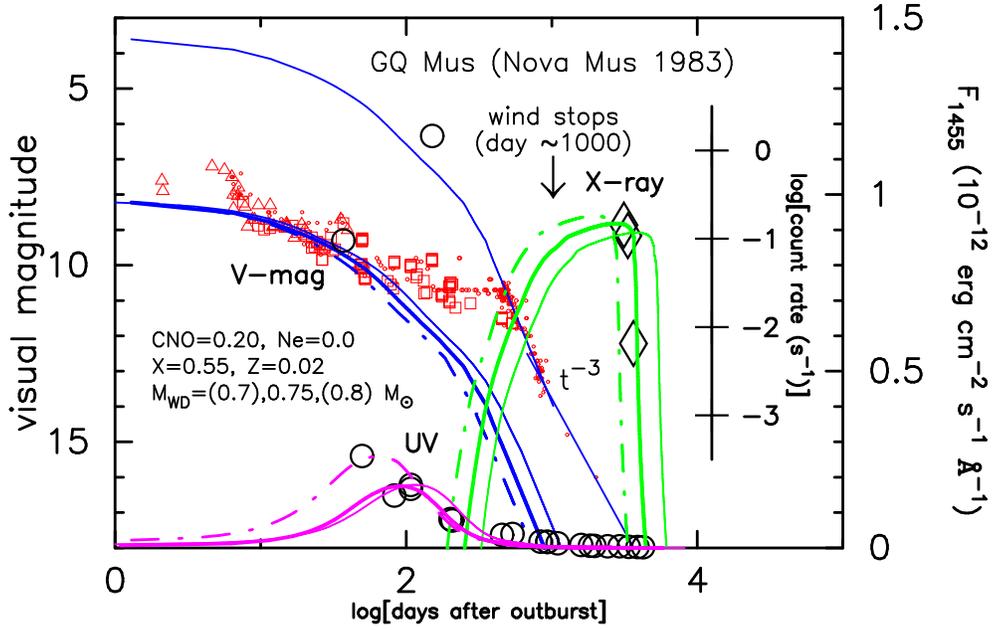}
\caption{
Calculated light curves of free-free $V$ magnitude,
UV 1455 \AA, and supersoft X-ray ($0.1-2.4$~keV) fluxes are
compared with the observations.  {\it Open diamonds}: {\it ROSAT}
X-ray (0.1--2.4 keV) observations from \citet{sha95} and
\citet{ori01}.  {\it Open circles}: {\it IUE} UV 1455 \AA\ 
observations.  {\it Open triangles}: visual and $V$ magnitudes
are from IAU Circulars 3764, 3766, 3771, 3782, and 3853.
{\it Open squares}: $V$ magnitude observations from \citet{whi84}
and {\it IUE} $V_{\rm FES}$ data.  {\it Small circles}: visual
magnitude observations from RASNZ and AAVSO.
Our best-fit model consists of the white dwarf with mass
$0.75~M_\sun$ for the envelope chemical composition of $X=0.55$,
$X_{\rm CNO} =0.20$, and $Z=0.02$.  An arrow indicates the epoch
when the optically thick wind stops.  The upper thin solid line
is the same as the thick solid line, but lifted up by 4.5 mag 
to represent contribution of strong emission lines to the $V$
bandpass (Papers I and II).  Model predictions for white
dwarf masses $M_{\rm WD}= 0.7 ~M_\sun$ ({\it thin solid line})
and $M_{\rm WD}= 0.8 ~M_\sun$ ({\it dash-dotted line}) are also
added for comparison.}
\label{all_mass_gq_mus_x55z02c10o10_compare}
\end{figure*}


\begin{figure*}
\epsscale{0.85}
\plotone{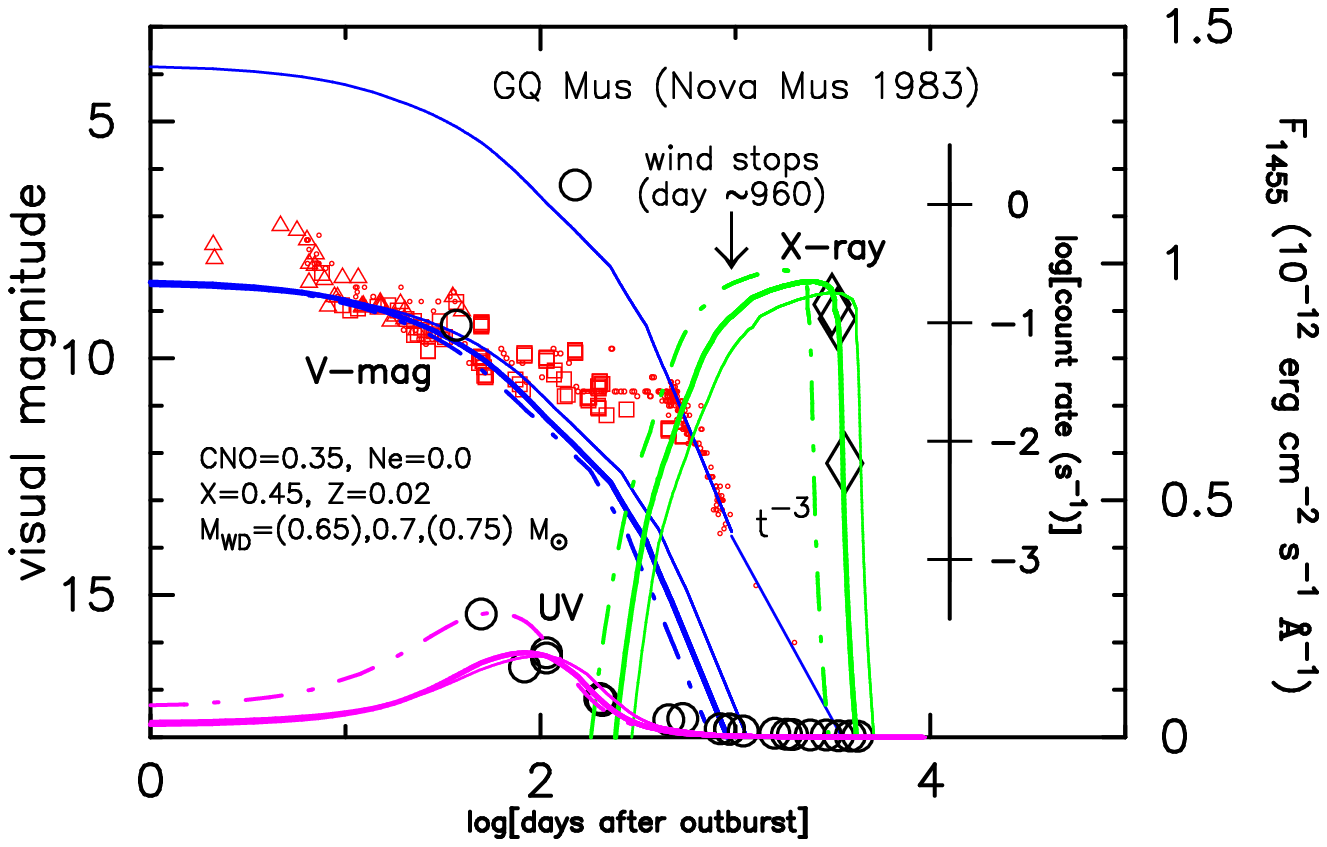}
\caption{
Same as in Fig. \ref{all_mass_gq_mus_x55z02c10o10_compare}, but for
our best-fit model of $M_{\rm WD}= 0.7 ~M_\sun$ for the
envelope chemical composition of
$X=0.45$, $X_{\rm CNO} =0.35$, and $Z=0.02$.
Two other cases of the white dwarf mass, $M_{\rm WD}= 0.65 ~M_\sun$
({\it thin solid line}) and $M_{\rm WD}= 0.75 ~M_\sun$ ({\it dash-dotted
line}) are also added for comparison.}
\label{all_mass_gq_mus_x45z02c15o20_compare_uv_x}
\end{figure*}


\begin{figure*}
\epsscale{0.85}
\plotone{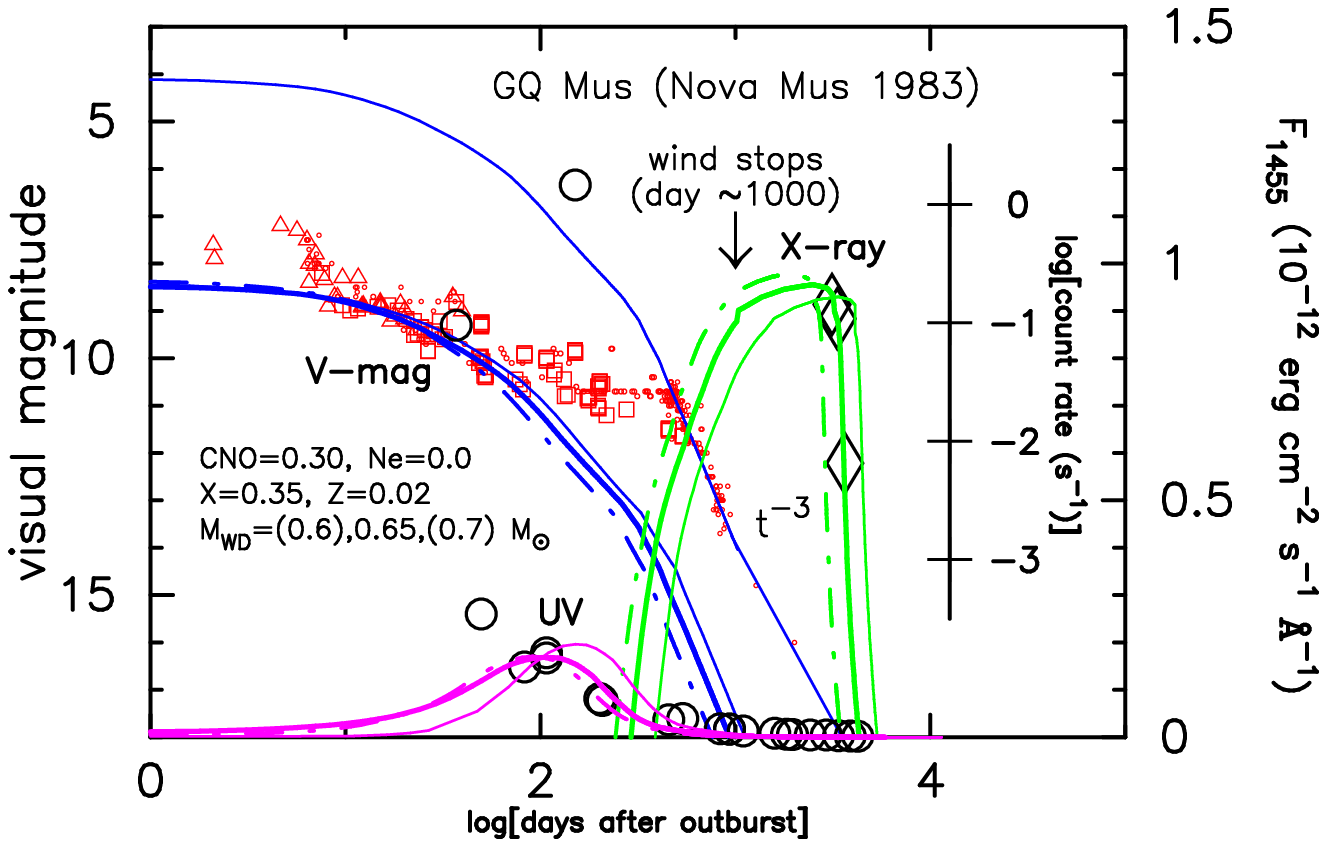}
\caption{
Same as Fig. \ref{all_mass_gq_mus_x55z02c10o10_compare},
but for our best-fit model of the white dwarf mass of $0.65~M_\sun$
for the envelope chemical composition of
$X=0.35$, $X_{\rm CNO} =0.30$, and $Z=0.02$.
Two other cases of the white dwarf mass, $M_{\rm WD}= 0.6 ~M_\sun$
({\it thin solid line}) and $M_{\rm WD}= 0.7 ~M_\sun$ ({\it dash-dotted
line}) are also added for comparison.}
\label{all_mass_gq_mus_x35z02c10o20_compare_uv_x}
\end{figure*}

\subsection{Multiwavelength light curves}
In the optically thick wind model, a large part of the envelope
is ejected continuously during a relatively long period
\citep[e.g.,][]{kat94h}.  After maximum expansion, the photospheric
radius gradually decreases keeping the total luminosity
($L_{\rm ph}$) almost constant.  The photospheric temperature
($T_{\rm ph}$) increases with time because of $L_{\rm ph} = 4 \pi
R_{\rm ph}^2 \sigma T_{\rm ph}^4$.  The maximum emission shifts
from the optical to supersoft X-ray through ultraviolet (UV) and
extreme ultraviolet (EUV).  This causes the luminosity to decrease
in the optical and to increase in the UV, until it reaches
a maximum. The following decay in the UV is accompanied
by an increase of the supersoft X-ray range.  These
timescales depend crucially on the WD parameters such as
the WD mass and the chemical composition of the envelope
\citep[e.g.][Paper I]{kat97}.  Thus, we can follow the development
of optical, UV, and supersoft X-ray light curves by a single
modeled sequence of steady wind solutions.

\subsection{Free-free light curves}
\label{free-free_light_curve}
Spectra of novae show blackbody features at very early stages,
but free-free emission from optically thin plasma will eventually
dominate \citep[e.g.,][]{gal76, kra84}.  During the optically
thick wind phase (see Paper I for more details), the extended
regions outside the photosphere are optically thin.  
The optical and IR free-free emission fluxes arising from these
regions can be estimated from 
\begin{equation}
F_\nu \propto \int N_e N_i d V
\propto \int_{R_{\rm ph}}^\infty {\dot M_{\rm wind}^2
\over {v_{\rm wind}^2 r^4}} r^2 dr
\propto {\dot M_{\rm wind}^2 \over {v_{\rm ph}^2 R_{\rm ph}}},
\label{free-free-wind}
\end{equation}
where $F_\nu$ is the flux at the frequency $\nu$,
$N_e$ and $N_i$ are the number densities of electrons
and ions, respectively, $V$ is the volume of the optically thin
region, $R_{\rm ph}$ is the photospheric radius,
$\dot M_{\rm wind}$ is the wind mass loss rate, and $v_{\rm ph}$
is the photospheric velocity.  In equation (\ref{free-free-wind})
we assume that the electron temperature and the degree of
ionization are constant in the free-free emitting region.
We also assume that $N_e \propto \rho_{\rm wind}$, $N_i \propto
\rho_{\rm wind}$, and use the continuity equation, i.e.,
$\rho_{\rm wind} = \dot M_{\rm wind}/ 4 \pi r^2 v_{\rm wind}$,
where $\rho_{\rm wind}$ and $v_{\rm wind}$ are the density and
velocity of the wind, respectively.  Finally, we assume
that $v_{\rm wind}= {\rm const.}= v_{\rm ph}$ in the optically 
thin region.

After the optically thick wind stops, the total mass of the
ejecta remains constant in time.  The flux from such homologously
expanding ejecta is approximately given by
\begin{equation}
F_\nu \propto \int N_e N_i d V
\propto \rho_{\rm ej}^2 V_{\rm ej} \propto \left({{M_{\rm ej}}
\over V_{\rm ej}} \right)^2 V_{\rm ej}~  \propto V_{\rm ej}^{-1}
 \propto  R^{-3}  \propto t^{-3},
\label{free-free-stop}
\end{equation}
where $\rho_{\rm ej}$, $V_{\rm ej}$, and $M_{\rm ej}
(= {\rm const.})$ are the density, volume, and total mass,
respectively, of the ejecta.  We assume that the ejecta are
expanding at a constant velocity, $v$.  So we have the radius of
the ejecta of $R = v t$, where $t$ is the time after the outburst.
The proportionality constants in equations (\ref{free-free-wind})
and (\ref{free-free-stop}) cannot be determined a priori
because radiative transfer is not calculated outside
the photosphere: these were determined using the procedure
described below in \S\ref{light_curve_fitting}.


\begin{figure*}
\epsscale{0.85}
\plotone{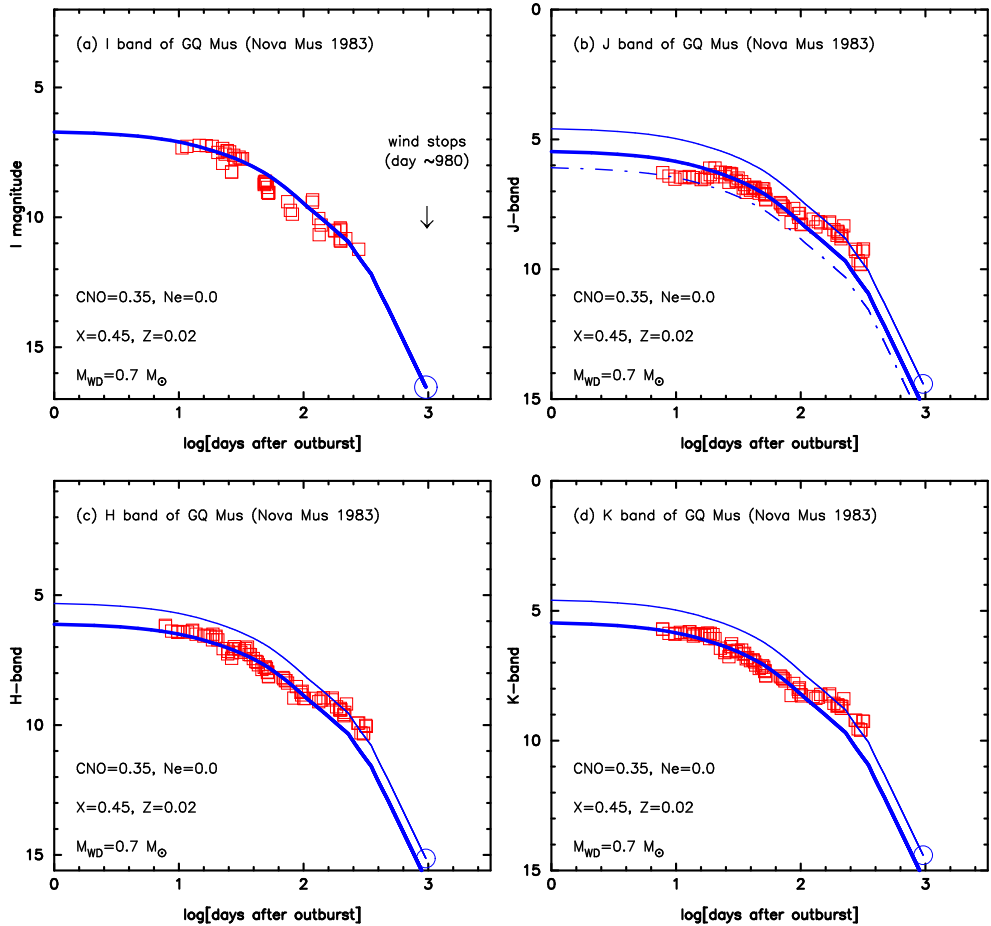}
\caption{
Near-infrared multiwavelength light curves ($I$, $J$, $H$, and
$K$) for free-free emission, based on equation (\ref{free-free-wind}),
and our model light curve corresponding to $M_{\rm WD} = 0.70
~M_\sun$, $X=0.45$, and $X_{\rm CNO}= 0.35$, together with the
observations.  {\it Thick solid line:} fit of the early phase.
{\it Thin solid line:} fit of the late phase after the transition
phase (see, e.g., Papers I and II).  {\it Open circles:} end of
the wind phase.  (a) $I$ magnitudes from free-free emission: These
thick and thin lines are overlapped.  (b) $J$ magnitudes of
free-free emission.  The \ion{He}{1}~$\lambda 10830$ contributes
to the $J$ band from the very early phase of the outburst
\citep{whi84, kra84}, so one more line ({\it dash-dotted}) is
added to represent the very early phase.  (c) $H$ magnitudes
of free-free emission.  (d) $K$ magnitudes of free-free emission.
Observational data ($I$, $J$, $H$, and $K$) are taken from 
\citet{whi84} and \citet{kra84}.  Essentially the same figures
are obtained for the other two cases of $M_{\rm WD} = 0.75 ~M_\sun$,
$X=0.55$, $X_{\rm CNO}= 0.20$ (see Fig. 
\ref{all_mass_gq_mus_x55z02c10o10_compare}), and $M_{\rm WD} = 0.65
~M_\sun$, $X=0.35$, and $X_{\rm CNO}= 0.30$ (see Fig.
  \ref{all_mass_gq_mus_x35z02c10o20_compare_uv_x}).}
  \label{gq_mus_m0700_x45z02c15o20_i_j_h_k}
\end{figure*}

\subsection{System parameters of optically thick wind model}
     The light curves of our optically thick wind model are
parameterized by the WD mass ($M_{\rm WD}$),
the chemical composition of the envelope ($X$, $X_{\rm CNO}$,
$X_{\rm Ne}$, and $Z$), and the envelope mass
($\Delta M_{\rm env, 0}$) at the time of the outburst (JD 2,445,348.0).
For the metal abundance we adopt $Z=0.02$,
which also includes carbon, nitrogen, oxygen, and neon
with solar composition ratios.

Three different sets of abundance determinations for the ejecta of
GQ Mus are available from the literature, but their values are
rather scattered, as it appears from Table
\ref{gqmus_chemical_abundance}.  We here consider three cases,
in order of decreasing hydrogen content in the envelope, not
directly corresponding to the three sets of abundance determination
in Table \ref{gqmus_chemical_abundance}:
\noindent a)  $X= 0.55$, $X_{\rm CNO}= 0.20$, $X_{\rm Ne}= 0.0$,
and $Z=0.02$;
\noindent b) $X= 0.45$, $X_{\rm CNO}= 0.35$, $X_{\rm Ne}= 0.0$,
and $Z=0.02$;
\noindent c) $X= 0.35$, $X_{\rm CNO}= 0.30$, $X_{\rm Ne}= 0.0$,
and $Z=0.02$. 
These composition sets correspond to assuming 25\%, 55\%,
and 100\% mixing of C+O WD matter with a hydrogen-rich envelope
with solar composition.  If we change $X_{\rm CNO}$ while
keeping the hydrogen content $X$ constant, the light curves
hardly change provided $X_{\rm CNO} \gtrsim 0.2$.

We have searched for the best fit model by changing the WD mass
in steps of $0.05~M_\sun$ for the above three sets of  chemical
compositions.

\section{Light curve fitting}
\label{light_curve_fitting}
We apply to GQ Muscae 1983 the model light curves of classical
novae described in the previous section , and evaluate its
fundamental parameters.  The optical $V$ magnitude and the near
infrared $I$, $J$, $H$, and $K$ magnitudes are calculated from
free-free emission, while the UV 1455 \AA\ and supersoft X-ray
fluxes are obtained from blackbody emission, as
illustrated in Figure \ref{sed_uv_opt_ir_filter}.

Models and observations are compared in Figures
\ref{all_mass_gq_mus_x55z02c10o10_compare} --
\ref{all_mass_gq_mus_x35z02c10o20_compare_uv_x} which refer to,
respectively, the three following chemical compositions sets:
$(X,Y,X_{\rm CNO},Z)=$ $(0.55, 0.23, 0.20, 0.02)$,
$(0.45, 0.18, 0.35, 0.02)$, and $(0.35, 0.33, 0.30, 0.02)$.
The white dwarf mass range explored is $M_{\rm WD}= 0.60$ to
$0.80 ~M_\sun$, in $0.05~M_\sun$ steps. In each figure,
the thick solid line represents the white dwarf mass value
that leads to the best representation of the data for the above
chemical composition sets.

\subsection{Supersoft X-ray fluxes}
The decay time of the supersoft X-ray flux is a good indicator
of the WD mass (Paper I).  Through a careful visual inspection
to the data, we have selected the models that best fitted
the {\it ROSAT} observations \citep{oge93, sha95, ori01, bal01}.
The model and the observed supersoft X-ray light curves
are reported in Figures \ref{all_mass_gq_mus_x55z02c10o10_compare}
-- \ref{all_mass_gq_mus_x35z02c10o20_compare_uv_x}.

\subsection{Continuum UV 1455 \AA~ fluxes}
The light curves in the UV 1455 \AA\ continuum are a good
indicator of photospheric temperature during the early decay
phase of novae.  Such curves are in general quite smooth,
as shown by \citet{cas02} for several objects observed with
{\it IUE}.  A comparison of models with observed UV light
curves (see Figs. \ref{all_mass_gq_mus_x55z02c10o10_compare} --
\ref{all_mass_gq_mus_x35z02c10o20_compare_uv_x}) shows that
this is also the case of GQ Mus, except for the two earliest
observations on day 37 and 49, and that of day 151.  Leaving
these data apart, we find that the model that best fits the data
is that of a white dwarf with mass $M_{\rm WD} = 0.75$, 0.7,
and $0.65~M_\sun$, for the three above chemical composition
sets, respectively.  As for the observation of day 151, the large deviation
from the model's smooth trend is clearly due to the occurrence of the UV flash
discussed in \S 3.2.  The otherwise good agreement between models and
observations shown in the above figures may suggest that also the observations
of day 37 and 49 are due to a UV flash (see discussion in 
\S\ref{UV_flushes}).


\begin{figure}
\epsscale{1.15}
\plotone{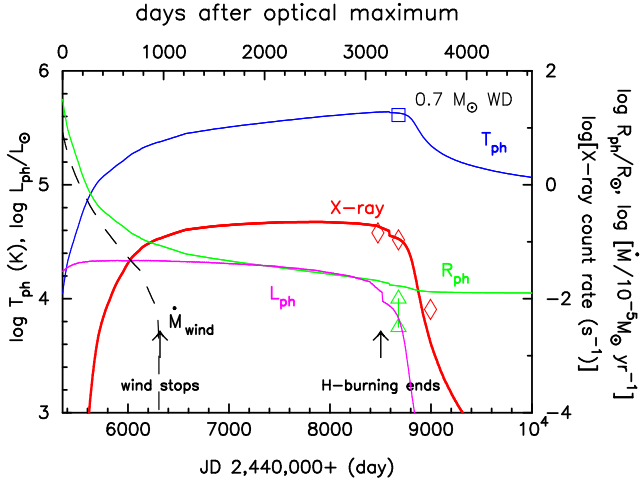}
\caption{
Evolution of wind mass loss rate ($\dot M_{\rm wind}$:
{\it dashed line}), photospheric temperature ($T_{\rm ph}$:
{\it upper thin solid line}), photospheric radius ($R_{\rm ph}$:
{\it middle thin solid line}), photospheric luminosity
($L_{\rm ph}$: {\it lower thin solid line}), and X-ray flux
({\it thick solid line}) of our $M_{\rm WD}= 0.7~M_\sun$ model
for the envelope chemical composition of $X=0.45$,
$X_{\rm CNO} =0.35$, and $Z=0.02$.  {\it Open diamonds}:
Observational X-ray count rates taken from \citet{sha95} and
\citet{ori01}.  {\it Open squares and open triangles}:
photospheric temperatures and photospheric radii, respectively,
taken from \citet{bal01}.
\label{gq_mus_softXray_r_t}}
\end{figure}

\subsection{Optical and infrared fluxes}
\label{optical_infrared}
The visual light curves computed from free-free emission are
compared with the observed ones in  Figures
\ref{all_mass_gq_mus_x55z02c10o10_compare} --
\ref{all_mass_gq_mus_x35z02c10o20_compare_uv_x} for the above
chemical composition sets.  As discussed in \S 1 and in Papers I
and II, visual magnitudes are contaminated by strong emission
lines which will eventually dominate over the continuum, causing
an increasing deviation from our free-free models.  In GQ Mus,
the forbidden [\ion{O}{3}] $\lambda\lambda$ 4959, 5007 emission
lines already appeared 39 days after the outburst \citep{kra84}.
At about this date the observed visual light curve did actually
start to show an increasing deviation from the free--free
emission model until, by about day 500, the contribution from
the emission lines stabilized so that the observed light curve
recovered the the shape of the model curve.  This effect is
approximatively taken into account by lifting up the template
light-curve of GQ Mus by about 4.5 mag, as shown in Figures
\ref{all_mass_gq_mus_x55z02c10o10_compare} --
\ref{all_mass_gq_mus_x35z02c10o20_compare_uv_x}.

On the other hand, the near infrared $IJHK$ bands are not so
heavily contaminated by emission lines, as shown in Figure
\ref{gq_mus_m0700_x45z02c15o20_i_j_h_k}, where the $IJHK$
photometric data from \citet{whi84} are compared with our
theoretical light curve.  The figure shows that the model
fits the observations reasonably well until about day 100
(thick solid line in the figure), being anyhow the deviations
after that date sensibly smaller than in the visual.
The contribution from emission lines is particularly small in 
the $I$, $H$, and $K$ bands, whereas the emission line of
\ion{He}{1} $\lambda 10830$ contributes somewhat to the $J$
band from the very early phase \citep{whi84, kra84}.
To mimic the different contribution by the emission lines at
different phases of the nova development, it is sufficient
to lift up the template free-free line two times (starting
from the dash-dotted line representing the very early phases,
to the thick solid line representing the intermediate phases,
and then finally to the thin solid line, representing the late
phases).


To summarize the results in this section, the white dwarf mass
that best reproduces the X-ray, UV, optical, and near IR is
$M_{\rm WD}= 0.7 \pm 0.05 ~M_\sun$, where the error bar reflects
mainly the uncertainty on chemical composition.

\section{Discussion}

\begin{figure}
\epsscale{1.15}
\plotone{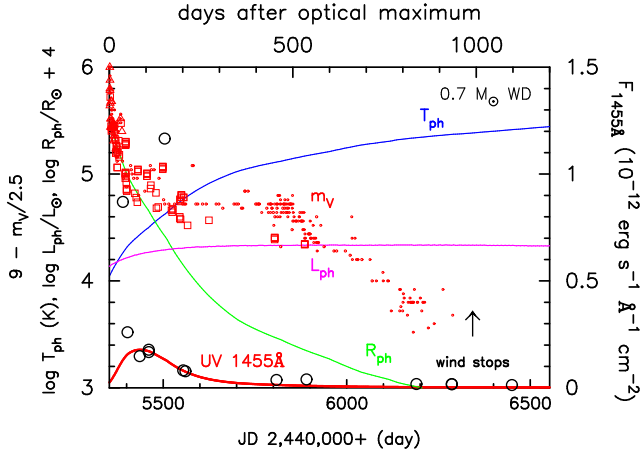}
\caption{
Photospheric temperature ($T_{\rm ph}$: {\it upper thin solid
line}), photospheric luminosity ($L_{\rm ph}$: {\it middle thin solid
line}), photospheric radius ($R_{\rm ph}$: {\it thin solid line from
left-upper to right-lower}), and UV 1455 \AA~ flux ({\it thick solid
line}) of our $M_{\rm WD}= 0.7~M_\sun$ model for the envelope chemical
composition of $X=0.45$ and $X_{\rm CNO} =0.35$.  {\it Open circles}: UV
1455 \AA~ fluxes.  Visual magnitude are also shown (as $9 - m_V /
2.5$) using the same symbols as in Figs.
\ref{all_mass_gq_mus_x55z02c10o10_compare} --
\ref{all_mass_gq_mus_x35z02c10o20_compare_uv_x}.
\label{gq_mus_uv1455_r_t}}
\end{figure}


\begin{figure}
\epsscale{1.15}
\plotone{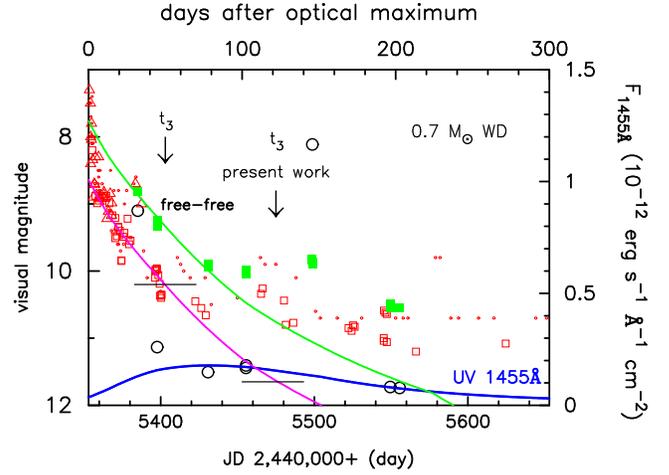}
\caption{ Visual and $V$ magnitudes during 300 days after the optical maximum.
The same symbols as in Figs.  \ref{all_mass_gq_mus_x55z02c10o10_compare} --
\ref{all_mass_gq_mus_x35z02c10o20_compare_uv_x} except the {\it IUE}
$V_{\rm FES}$ magnitude ({\it filled squares}).
Two free-free light curves are plotted both
for the {\it IUE} $V_{\rm FES}$ magnitudes and for Whitelock et
al.'s (1984) $V$ magnitudes.  The $V_{\rm FES}$ data are $0.5-0.8$ mag
brighter than those of Whitelock et al.'s.
The $t_3$ time ({\it left arrow}) is estimated to be 50 days from
Whitelock et al.'s $V$ magnitudes together with the peak brightness of
7.2 mag \citep{lil83}.  Note that the decay time
({\it right arrow}: $t_3= 122$ days)
corresponding to our universal decline law 
({\it lower solid free-free line}) is sensibly larger
than that previously quoted.   See text for more details.
\label{gq_mus_t3_uv1455_m0700_x45z02c15o20}}
\end{figure}


\begin{figure}
\epsscale{0.9}
\plotone{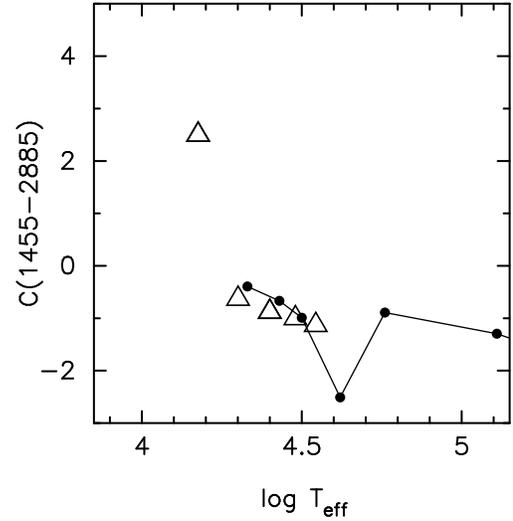}
\caption{
Color-temperature relation for nova wind solutions.
The UV color of $C(1455-2885)$ is defined by
$-2.5 \log( F(1455) / F(2885))$.
{\it Filled circles connected with a solid line}:
reddening-corrected $C(1455-2885)$ v.s.
the photospheric temperature $T_{\rm eff}$
corresponding to our best fit model.
{\it Open triangles}: color-temperature relations estimated from
Fig. 10 of \citet{hau97} and Fig.8 of \citet{sho99}.
See text for more details.
\label{uv_outburst}}
\end{figure}


\begin{figure}
\epsscale{0.9}
\plotone{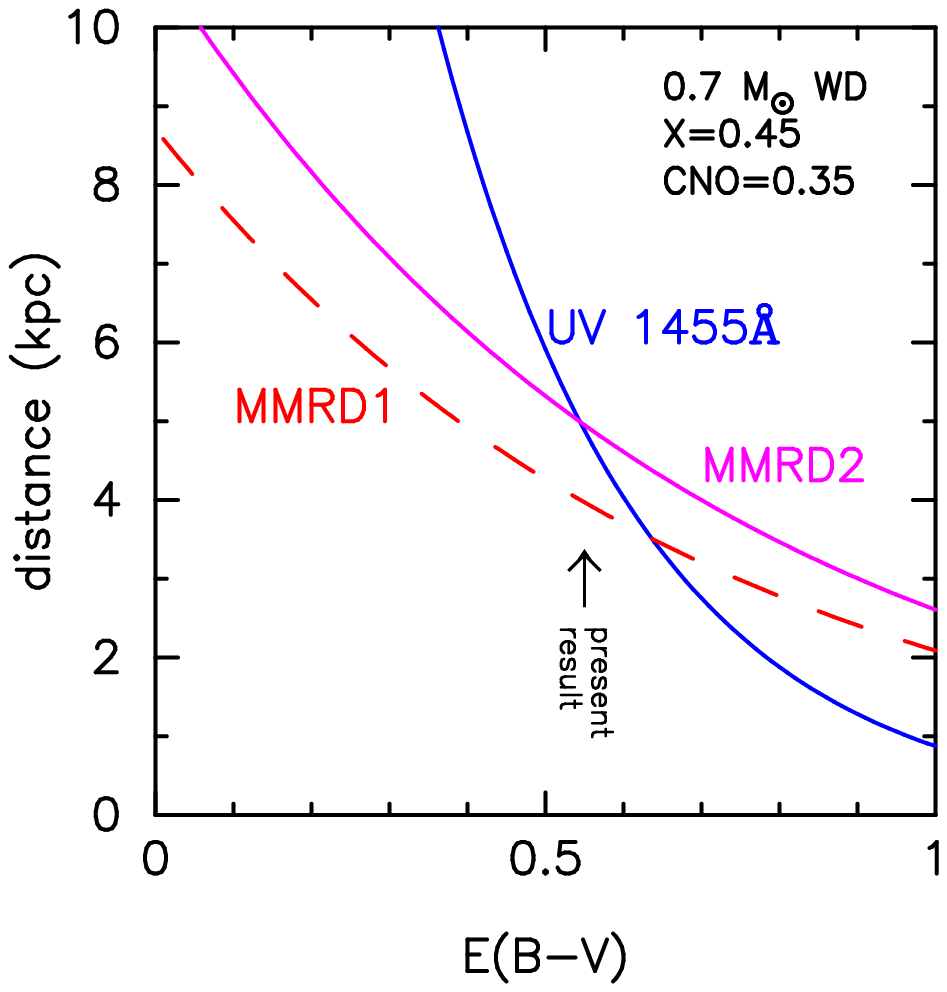}
\caption{Distance-reddening relations derived from the UV 1455 \AA\  
fitting (solid line labeled ``UV 1455 \AA\ '')
of our model $M_{\rm WD}= 0.7 ~M_\sun$, $X=0.45$,
and $X_{\rm CNO}= 0.35$ and the maximum magnitude vs. rate of decline
(labeled ``MMRD1'' and ``MMRD2'').  The dashed line labeled 
``MMRD1'' is a MMRD relation calculated from Schmidt-Kaler's law, i.e.,
eq. (\ref{schmidt-kaler-law}) together with $t_3 = 50$ days.
The solid line labeled ``MMRD2'' is the
same MMRD relation calculated from $t_3 = 122$ days estimated from our
universal decline law.  An arrow indicates our value
of $E(B-V)= 0.55$, which is an error-weighted mean of
$E(B-V)=0.51 \pm 0.06$ and $0.58 \pm 0.04$ in \S 3.1.
See text for more details.
\label{gq_mus_distance_reddening_m0700_x45z02c15o20}}
\end{figure}

\subsection{Overall  Development of the Nova Outburst}
In this subsection, we analyze the overall development of
the nova outburst based on our model of $M_{\rm WD}= 0.7 
~M_\sun$, $X=0.45$, $X_{\rm CNO}= 0.35$, and $Z=0.02$
(Fig. \ref{all_mass_gq_mus_x45z02c15o20_compare_uv_x}).
Figures \ref{gq_mus_softXray_r_t} -- 
\ref{gq_mus_t3_uv1455_m0700_x45z02c15o20} report some relevant
model predictions and observations as a function of time.
First we point out two important epochs of the nova development:
one is the end of optically thick wind phase on day 1000, and
the other is the end of hydrogen shell burning on day 3300
as summarized in Table \ref{properties_gq_mus}.  Because of
the rapid shrinking of the the photospheric radius ($R_{\rm ph}$),
also the photospheric temperature ($T_{\rm ph}$) rapidly
increases until the wind stops, as indicated in Figure 
\ref{gq_mus_softXray_r_t}.  Also the wind mass loss rate
($\dot M_{\rm wind}$) decreases rapidly from $\sim 2 \times
10^{-4} M_\sun$~yr$^{-1}$ to $1 \times 10^{-7} M_\sun$~yr$^{-1}$
during the first 1000 days.  Thus the total mass ejected by the
winds amounts to $\sim 2 \times 10^{-5} M_\sun$.  After the wind
stops, the photospheric radius becomes smaller than $\sim 0.1
~R_\sun$, and the photospheric temperature eventually becomes
larger than 20 eV, so that the nova enters a supersoft X-ray
phase.  After the hydrogen shell burning ends, the nova cools
down rapidly, and a fast decay phase of the supersoft X-ray flux
follows, as already shown in the previous section.  In what
follows, we summarize some relevant observational aspects and
make a comparison with our model predictions.

From the coronal lines in the optical spectrum \citet{dia92}
deduced a luminosity of $10^{37} - 10^{38}$ ergs~s$^{-1}$
and a temperature of $(2-3) \times 10^5$~K.
These values are consistent with our model of a $0.7~M_\sun$
WD, which provides $T_{\rm ph} = 4 \times 10^5$~K and
$L_{\rm ph}= 6 \times 10^{37}$ ergs~s$^{-1}$ about 2700 days
after the outburst, as it may be appreciated from Figure
\ref{gq_mus_softXray_r_t}.

\citet{dia95} estimated a central source temperature of
164,000~K, a luminosity of $350 ~L_\sun$, and an ejecta mass
of $5 \times 10^{-4} M_\sun$, about 4000 days after the outburst.
For our best-fit model of a  $0.7 ~M_\sun$ white dwarf
with an initial envelope mass of $\Delta M_0 = 2.7 \times 10^{-5}
M_\sun$, the total mass lost by the optically thick wind is
$\Delta M_{\rm wind} = 1.9 \times 10^{-5} M_\sun$.
We also obtain $T_{\rm ph} = 1.5 \times 10^5$~K
and $L_{\rm ph}= 4 \times 10^{35}$ergs~s$^{-1} \approx 100
~L_\sun$, 4000 days after the outburst, as shown in Figure
\ref{gq_mus_softXray_r_t}.   Our values are then very consistent
with those obtained by \citet{dia95}, except for their larger
ejected mass (see below for another estimate).

The envelope mass itself depends mainly on the white dwarf
mass and slightly on the chemical composition.
For a $0.65 ~M_\sun$ white dwarf, $X=0.35$,
and $X_{\rm CNO}= 0.30$, we obtain a slightly larger
envelope mass, $\Delta M_0 = 3.6 \times 10^{-5} M_\sun$,
and a larger wind mass loss, $\Delta M_{\rm wind} = 
2.3 \times 10^{-5} M_\sun$.  On the other hand, for a
$0.75 ~M_\sun$ white dwarf with $X=0.55$, and $X_{\rm CNO}= 0.20$,
we obtain a slightly smaller envelope mass, $\Delta M_0 = 2.6 
\times 10^{-5} M_\sun$, and a smaller wind mass loss, $\Delta
M_{\rm wind} = 1.7 \times 10^{-5} M_\sun$.

Using their photoionization model, \citet{mor96b} estimated the
temperatures during the static phase and at the hydrogen
burning turnoff to be $2.5 \times 10^5$ K and $4.1 \times
10^5$ K, respectively, and an ejected shell mass of $8 \times
10^{-5} M_\sun$.  These values are consistent with our model,
i.e., $> 2.4 \times 10^5$ K, $> 4.4 \times 10^5$ K, and $2 
\times 10^{-5} M_\sun$.   The above authors determined the
duration of the wind phase and of the static hydrogen burning
phase to be $< 0.52$ yr and $8.8$ yr, to be compared with our
values of 2.7 yr and 6.3 yr, respectively.   The much longer
duration of the wind phase obtained by us is likely due to
the rapid decrease of the wind mass loss rate with an e-folding
time of 0.24 yr, corresponding to a decrease by a factor of
10 in 0.7 yr.

\subsection{Photospheric Temperature Development} 
Figure \ref{uv_outburst} shows the reddening corrected UV color
index $C(1455-2885)$ as a function of the photospheric temperature
obtained from our best-fit model at the time of the individual
observations (filled circles).  The figure indicates that the
color index decreases smoothly with increasing temperature
except for its abrupt decrease at the time of the UV flash on
day 151 (note that the temperature increases with time in these
early phases).

It is interesting to compare the color--$T_{\rm ph}$ values reported
in Figure \ref{uv_outburst} with those derived from \citet{hau97}
and \citet{sho99}, who calculated early-time nova spectra using
non-LTE atmosphere codes with winds having density and velocity laws
within the nova envelope.  For this comparison, we have taken
the values of $C(1455-2885)$ from Figure 10 of \citet{hau97} for
$T_{\rm eff}= 15,000$, 20,000, 25,000, and 30,000 K, and from
Figure 8 of \citet{sho99} for $T_{\rm eff}= 35,000$ K.
The model values of $C(1455-2885)$
so obtained are plotted as a function of the corresponding
effective temperature in Figure \ref{uv_outburst} (open triangles).
It clearly appears from Figure \ref{uv_outburst} that the two sets
of values are very consistent with each other, except for the UV
flash on day 151.  This result supports our assumption that the
UV 1455 \AA\ flux is reasonably well accounted for by blackbody
emission.  Indeed, line blanketing is rather small at this
wavelength, so that the blackbody model does not deviate much
from Hauschildt et al.'s non-LTE models, as already discussed
in Paper I.

\subsection{Emergence of the secondary component}
The mass of the donor star (the secondary component) can be
estimated from the orbital period.  \citet{dia89, dia94}
obtained $P_{\rm orb}= 0.05936$~days ($1.425$~hr) from the
orbital modulations with an amplitude of 0.6 mag.  Using
Warner's (1995) empirical formula
\begin{equation}
{{M_2} \over {M_\sun}} \approx 0.065 \left({{P_{\rm orb}}
\over {\rm hours}}\right)^{5/4},
\mbox{~for~} 1.3 < {{P_{\rm orb}} \over {\rm hours}} < 9
\label{warner_mass_formula}
\end{equation}
we get $M_2 = 0.10 ~M_\sun$.  The orbital separation is then
$a = 0.59 ~R_\sun$ for $M_{\rm WD}= 0.7 ~M_\sun$, the effective
radius of the Roche lobe for the primary component (white dwarf)
is $R_1^* = 0.33 ~R_\sun$, and the effective radius of the
secondary is $R_2^* = 0.14 ~R_\sun$.  In our model, the companion
emerges from the white dwarf envelope when the photospheric
radius of the white dwarf shrinks to $R_{\rm ph} \sim 0.6 ~R_\sun$
(the separation) or $0.3~R_\sun$ (the Roche lobe).  This 
happens on about day~330 or 500, respectively, in our best-fit
model with $M_{\rm WD}= 0.7 ~M_\sun$, as shown in Figures 
\ref{gq_mus_softXray_r_t} and \ref{gq_mus_uv1455_r_t}.



According to \citet{whi84}, strong infrared coronal lines appeared
57 days after optical maximum in V1500 Cyg, but were not present
in GQ Mus as late as 97 days after maximum.  Their appearance
roughly coincides with the emergence of the companion from the
white dwarf photosphere: this suggests that these strong coronal
lines arise from the shock between the white dwarf wind and
companion star.  Indeed, the emergence of the companion in V1500
Cyg took place about 50 days after optical maximum (Paper I),
which is consistent with the appearance of the line 57 days after
maximum.  In the case of GQ Mus, the absence of strong coronal
lines on day 97 days is consistent with our model in which the
companion emerged about 330 or 500 days after maximum.

In their study of GQ Mus, \citet{kra89} reported that the
[\ion{Fe}{10}] $\lambda 6374$ coronal line first appeared 2.2 to
3.4 yrs after the outburst, and became the strongest about 4 years
after it.  They argued that the coronal line was due to
photoionization from a hot radiation source rather than to
collisional excitation.   In our $0.7 ~M_\sun$ white dwarf model,
the optically thick wind stopped about 960 days (2.7 yrs) after
the outburst and then the photospheric temperature gradually
increased to above $20-25$ eV 4 yrs after the outburst, as shown
in Figures \ref{gq_mus_softXray_r_t} and \ref{gq_mus_uv1455_r_t}.
These authors suggested that $\sim 3 \times 10^5$ K in 1984,
which is consistent with our photospheric temperature of
$T_{\rm ph}= 2.2 \times 10^5$~K 2.2 yr after the outburst
(Figs. \ref{gq_mus_softXray_r_t} and \ref{gq_mus_uv1455_r_t}).

\subsection{Hard X-ray component}
\citet{oge87} reported that the count rate of the {\it EXOSAT}
low-energy telescope was about $3 \times 10^{-3}$ c~s$^{-1}$
400 days to 700 days after optical maximum, a value that
gradually decreased to $1 \times 10^{-3}$ c~s$^{-1}$ about
900 days after it.

If the X-rays originates from the shock between the wind and
the companion star, the soft X-ray behavior is consistent with
our $0.7 ~M_\sun$ white dwarf model (as well as $0.75 ~M_\sun$
and $0.65 ~M_\sun$ white dwarf models), because the emergence
of the companion star is predicted to happen on day 330, and the
wind stopped at 960 days after optical maximum.


\begin{deluxetable*}{lllll}
\tabletypesize{\scriptsize}
\tablecaption{Properties of GQ Muscae
\label{system_parameters}}
\tablewidth{0pt}
\tablehead{
\colhead{subject} &
\colhead{...} &
\colhead{data} &
\colhead{units} &
\colhead{reference}
}
\startdata
discovery & ... & 2,445,352.6 & JD &  \citet{lil83} \\
nova speed class & ... & moderately fast & & \citet{whi84} \\
$t_2$ & ... & 18 & days &  \citet{whi84}  \\
$t_3$ & ... & 48 & days & \citet{whi84} \\
$M_{V, {\rm peak}}$ from $t_3$ &... & $-7.55$ & mag & \citet{whi84} \\
distance from $t_3$ &... & 5 & kpc & \citet{whi84} \\
dust & ... & no & & \citet{whi84} \\
orbital period & ... & 1.425 & hr & \citet{dia95} \\
$E(B-V)$ & ... & 4.5 & & \citet{kra84} \\
wind phase & ... & $< 0.5$ & yr & \citet{mor96b} \\
H-burning phase & ... & 3300 & days & \citet{sha95}
\enddata
\end{deluxetable*}


\begin{deluxetable}{llll}
\tabletypesize{\scriptsize}
\tablecaption{Summary of our model
\label{properties_gq_mus}}
\tablewidth{0pt}
\tablehead{
\colhead{subject} &
\colhead{...} &
\colhead{data} &
\colhead{units}
}
\startdata
outburst day & ... & 2,445,348.0 & JD \\
opt. maximum & ... & 2,445,352.6 & JD  \\
$t_2$ & ... & 67\tablenotemark{a} & days  \\
$t_3$ & ... & 122\tablenotemark{a} & days \\
$M_{V, {\rm peak}}$ from $t_3$ &... & $-6.53$\tablenotemark{b} & mag \\
distance from $t_3$ &... & $5.0 \pm 0.3$ & kpc \\
secondary mass & ... & 0.1\tablenotemark{c} & $M_\sun$ \\
$E(B-V)$ & ... & $0.55 \pm 0.05$ & \\
distance by UV fit & ... & $4.9 \pm 0.9$  & kpc \\
$t_{\rm break}$\tablenotemark{d} & ... & 275 & day \\
WD mass & ... & $0.7 \pm 0.05$ & $M_\sun$ \\
WD envelope mass & ... & $(2.6 - 3.6) \times 10^{-5}$ & $M_\sun$ \\
mass lost by wind & ... & $(1.7 - 2.3) \times 10^{-5}$ & $M_\sun$ \\
wind phase & ... & $1000$ & days \\
H-burning phase & ... & $3300$ & days \\
separation & ... & 0.6 & $R_\sun$ \\
companion's emergence & ... & $330 \pm 30$ & days \\
UV flash luminosity & ... & $\gtrsim 45,000$\tablenotemark{e} & $L_\sun$
\enddata
\tablenotetext{a}{$t_2$ and $t_3$ are calculated from our fitted
universal decline law.}
\tablenotetext{b}{calculated from equation (\ref{schmidt-kaler-law})
together with $t_3$ above.}
\tablenotetext{c}{estimated from equation (\ref{warner_mass_formula}).}
\tablenotetext{d}{see Paper I.}
\tablenotetext{e}{estimated from a blackbody of $T \sim 10^5$~K.}
\end{deluxetable}

\subsection{Distance}
\citet{whi84} and \citet{kra84} have estimated the distance to
GQ Mus from the apparent and the absolute $V$ magnitudes at
maximum, $m_{\rm V}$ and $M_{\rm V}$.  This latter was computed
from the $t_3$ time through the Schmidt--Kaler Maximum Magnitude
Rate of Decline (MMRD) relation \citep{sch57}:
\begin{equation}
M_V = -11.75 + 2.5 \log t_3.
\label{schmidt-kaler-law}
\end{equation}
Despite the sensibly different input values adopted of 48 and 40
days for $t_3$, 0.4 and 0.45 for $E(B-V)$, and 7.2 and 6 for
$m_V$, these authors obtained quite similar values for the
distance, 5 kpc and 4.8 kpc, respectively.

A re-analysis of the merged visual magnitude data present in the
literature (see plot in Fig. 
\ref{gq_mus_t3_uv1455_m0700_x45z02c15o20}) confirms $t_3
\approx 50$ days, in close agreement with \citet{whi84},
so that $M_V =-7.5$ [cf. equation (\ref{schmidt-kaler-law})].
If, together with \citet{whi84}, we take $m_{V, {\rm max}} =
7.2$ as the best estimate of the apparent magnitude at optical
maximum \citep{lil83}, we finally obtain the following relation
between distance and $E(B-V)$:
\begin{equation}
m_V- M_V = -5 ~+~ 5 \log~d ~+~ 3.1 E(B-V)= 14.7,
\label{modulus}
\end{equation}
which is labeled ``MMRD1'' in Figure
\ref{gq_mus_distance_reddening_m0700_x45z02c15o20}.
In particular, the value $E(B-V)= 0.55 \pm 0.05$ obtained in
\S 3.1, corresponds to a distance of $d = 4.0 \pm 0.3$~kpc. 

Following the same procedure outlined in Paper I and
\citet{kat05h, kat07h}, an independent estimate of the distance
to GQ Mus can be obtained by comparing the observed light curve
in the 1455 \AA\ continuum with the corresponding model fluxes
in Figures \ref{all_mass_gq_mus_x55z02c10o10_compare} --
\ref{all_mass_gq_mus_x35z02c10o20_compare_uv_x},
\ref{gq_mus_uv1455_r_t},
and \ref{gq_mus_t3_uv1455_m0700_x45z02c15o20}.
The calculated flux at $\lambda$ = 1455 \AA\ at a distance of
10~kpc for our adopted model ($0.7 ~M_\sun$, $X=0.45$,
$X_{\rm CNO}= 0.35$) is $F_{\lambda}^{\rm mod}$ = 2.85 $\times
10^{-12}$~ergs~cm$^{-2}$~s$^{-1}$~\AA$^{-1}$ on JD 2,445,455.6.
The observed flux at the same date is $F_{\lambda}^{\rm obs}$=
1.78 $\times 10^{-13}$~ergs~cm$^{-2}$~s$^{-1}$ ~\AA$^{-1}$.
From these values one obtains the following relation between
distance and reddening:
\begin{equation}
  m_{\lambda}^{\rm obs} - M_{\lambda}^{\rm mod}=
5 \log ({d \over {10\mbox{~kpc}}}) + A_{\lambda} E(B-V) = 3.01,
\label{eq:ebv}
\end{equation}
where $m_{\lambda} = -2.5 \log(F_{\lambda})$, and $A_{\lambda}=
8.3$ \citep{sea79}.  This curve, labeled ``UV 1455 \AA,''
is reported in Figure 
\ref{gq_mus_distance_reddening_m0700_x45z02c15o20}.
In particular, equation (\ref{eq:ebv}) provides a distance of
$4.9 \pm 0.9$~kpc for $E(B-V)=0.55 \pm 0.05$, which is sensibly
larger than that obtained from MMRD1.


\begin{figure}
\epsscale{1.0}
\plotone{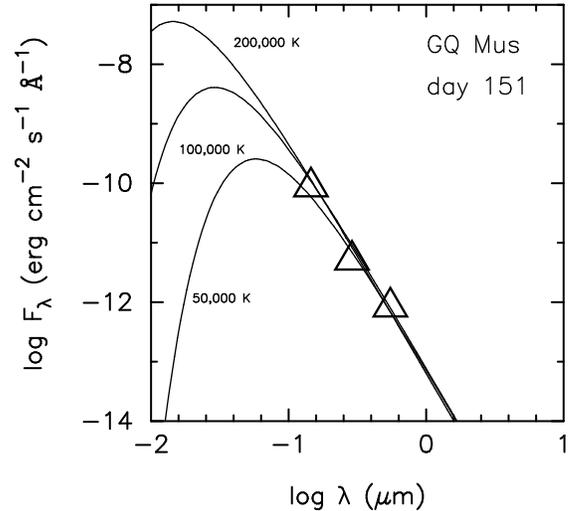}
\caption{
Energy distribution for three bands of UV 1455 \AA, 2885 \AA, and
{\it IUE} FES visual ({\it open triangles}) on day 151.
Blackbody energy distributions are also plotted for three temperatures
of 50,000, 100,000, and 200,000 K.
\label{sed_uv_outburst}}
\end{figure}

An alternative way to determine the distance of  GQ Mus is
to compare the shape of the theoretical visual light curve
(based on the free--free model) with the
observed visual fluxes. Such a comparison, done in Figure
\ref{all_mass_gq_mus_x45z02c15o20_compare_uv_x} is very
instructive because it shows that the object was up to 1.5
magnitudes brighter than predicted by our universal light
curve model during the earliest phases (until day 10). It is
then legitimate to consider GQ Mus as a super--bright nova.
An extreme case of this type was V1500 Cyg, which,
near peak luminosity was about 4 magnitudes
brighter than the Eddington limit for
a 1.0 $M_\sun$ white dwarf \citep{fer86}.
It is interesting to note that if such a super--bright phase
is ignored in GQ Mus, the $t_3$ and $t_2$ times
of the theoretical light curve that best fits the observations
are considerably longer than the observation indicates at
face value: we in fact estimate $t_3$ = 122 days and $t_2$= 67
days.  From equation (\ref{schmidt-kaler-law}) we then obtain
$M_{\rm ff, max} = -6.53$ at the time of maximum light ($t =
8$ days).  Since the apparent magnitude at visual maximum of
our theoretical light curve is $m_{\rm ff, max} = 8.66$ mag 
(see Fig. \ref{all_mass_gq_mus_x45z02c15o20_compare_uv_x}),
the distance to GQ Mus can be estimated to be $5.0 \pm 0.3$~kpc
for $E(B-V) = 0.55 \pm 0.05$.   If the reddening is not fixed,
the following relation applies:
\begin{equation}
m_{\rm ff, max} - M_{\rm ff, max}= - 5~+~5 \log d~+~3.1 E(B-V)= 15.19 ,
\label{mmrd_relation_2}
\end{equation}
which is plotted in Figure
\ref{gq_mus_distance_reddening_m0700_x45z02c15o20}
(labeled ``MMRD2'').
The two curves UV 1455 \AA\  and MMRD1 cross at $d= 3.5$ kpc
and $E(B-V)= 0.64$, whereas the curves UV 1455 \AA\  and MMRD2
cross at $d= 5.1$ kpc and $E(B-V)= 0.54$, as it appears from
Figure \ref{gq_mus_distance_reddening_m0700_x45z02c15o20}.
Since $E(B-V)= 0.55$ is the weighted mean of the reddening
determinations done in \S 3.1, we take this value together with
$d$= 5 kpc as the best estimates compatible with the observations.
The errors on the reddening and the distance are $\approx$ 0.05
dex and $\approx$ 0.5 kpc, respectively.

\subsection{The UV flashes}
\label{UV_flushes}
Here we estimate the total luminosity of the UV flash
on day 151 and discuss its outburst nature.  From Figure 
\ref{plotcont}, we have obtained the excess of energy above the
smooth decline for each of the three bands, i.e., 1455 \AA,
2885 \AA, and $V_{\rm FES}$, which are plotted in Figure 
\ref{sed_uv_outburst}.  These fluxes are consistent with
the energy distribution from a blackbody with a temperature of
$\gtrsim 100,000$~K.  \citet{has90} also estimated the temperature
from a Zanstra-like method, based on the \ion{He}{2}$/$H$\beta$
ratio, to be 85,000 to 100,000~K.  The photospheric temperature
of our $0.7 ~M_\sun$ white dwarf model are about 30,000~K (on day
108) before the UV flash and about 60,000~K (on day 202) after
the UV flash.

Assuming blackbody emission, we can estimate the size of the
emitting region from $(R/d)^2 \approx 10^{-21}$ for $T =
100,000$ K.  Taking $d = 5$ kpc, the radius of the emission region
is $R \sim 7 ~R_\sun$.  The total flux is estimated to be $1.7 
\times 10^{38}$ erg~s$^{-1}$ ($\approx 45,000 ~L_\sun$) from 
$L = 4 \pi R^2 \sigma T^4$.  On day 151, the photospheric radius
of our $0.7 ~M_\sun$ white dwarf model is as small as $1.3
~R_\sun$ and the photospheric temperature is as low as 42,000 K.
Therefore, the estimated radius and blackbody temperature,
$R \sim 7 ~R_\sun$ and $T \gtrsim 100,000$~K, suggest an episodic
expansion and strong heating of the photosphere.  This
was probably due to a dynamical mass ejection episode and to
strong shock heating.  The UV flash is anyhow a short-lived
event that hardly contributes to the total ejecta mass, as
suggested by the fast recovery of the emission line spectrum
by day 202 and by the negligible effect on the optical
light curves.
 
The first observation of {\it IUE} on day 37 also shows a high
UV flux compared with our white dwarf model.  A close look at
the optical brightness in Figure 
\ref{gq_mus_t3_uv1455_m0700_x45z02c15o20} indicates a small peak
around day 37.  This event may also be due to another mass
ejection episode.  Note that these mass ejection episodes add to
the major underlying continuous mass ejection process from the
optically thick wind, which endures until day $\sim 1000$, as
shown in Figure \ref{gq_mus_softXray_r_t}.

\section{Conclusions}
We have applied the ``universal decline law'' of classical novae
described in \S\ref{model_nova_outburst} to GQ Muscae 1983
and derived various parameters of the nova.
Our main results may be summarized as follows:

1. We  show that the ``universal decline law'' reproduces well
the observed light curves of GQ Mus in the optical
and in the near infrared $I$, $J$, $H$, and $K$ bands.

2. Our blackbody light curve model for the UV 1455 \AA\ band
can reproduce the observed UV 1455 \AA\ fluxes except two UV
flashes on day 37 and 151. 

3. The UV flash on day 151 described in \S\ref{uv_evolution_cont}
was accompanied by a mass ejection episode seen as a fast wind
with a terminal velocity of about 3200 km~s$^{-1}$.

4. An analysis of the {\it IUE} reprocessed data of GQ Mus
indicates $E(B-V)= 0.55 \pm 0.05$, a value that is larger
than previously reported.

5. We find that the mass of the WD component of GQ Mus is $0.7
\pm 0.05 ~M_\sun$ for the adopted envelope chemical composition
of $X=0.35-0.55$, $X_{\rm CNO}= 0.2-0.35$.  This value has been
derived by comparing predicted fluxes from our models with the
observations in the supersoft X-ray, in the UV, and in the
optical and near infrared, obtained at different times.

6. We have estimated a mass of $\Delta M_{\rm wind} \sim 2
\times 10^{-5} M_\odot$ lost by the wind.

7. We have estimated a distance of $d \sim 5$ kpc.



\acknowledgments We thank Albert Jones (RASNZ) for providing us
with their machine readable visual estimates of GQ Mus 1983 and
also AAVSO for the visual data of GQ Mus 1983.  We are also grateful
to the anonymous referee for useful comments that helped to improve
the paper.  This research has been supported in part by the
Grant-in-Aid for Scientific Research (16540211, 16540219, 20540227)
of the Japan Society for the Promotion of Science and by INAF PRIN 2007.





\appendix


\begin{figure}
\epsscale{0.8}
\plotone{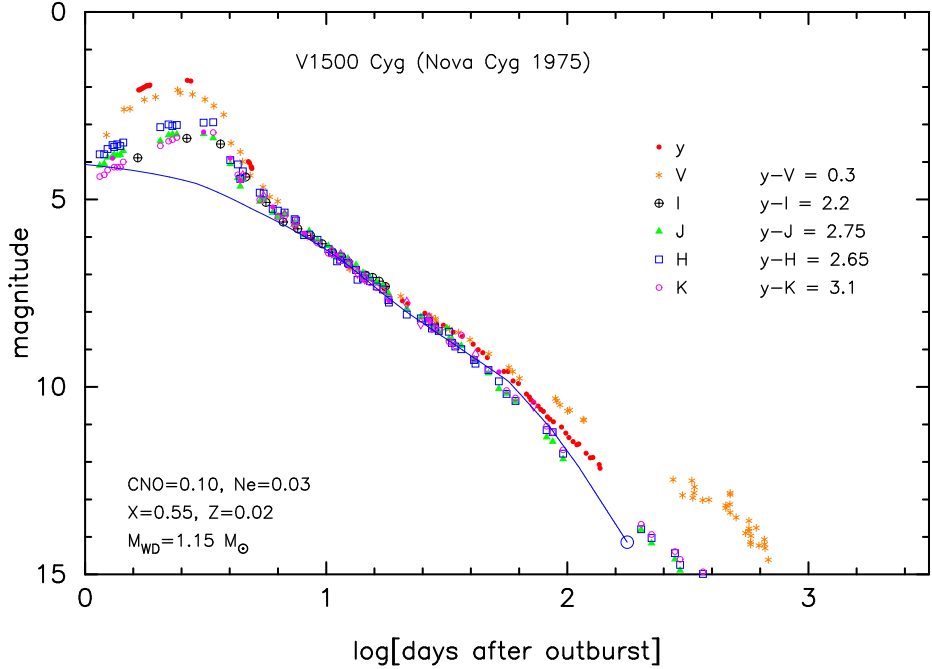}
\caption{
Observed $y$, $V$, $I$, $J$, $H$, and $K$ light curves for the classical
nova V1500 Cyg.  Each light curve is shifted down to the $y$ light
curve by $\Delta m_\lambda = \Delta V= y-V=0.3$, $\Delta I= y-I=2.2$,
$\Delta J= y-J=2.75$, $\Delta H= y-H=2.65$, and $\Delta K= y-K=3.1$.
These light curves are well overlapped between day 6 and 30 (days
after outburst).  Filled circles: $y$ magnitudes. 
Asterisks: $V$ magnitudes.  Plusing open circles: $I$ magnitudes.
Filled triangles: $J$ magnitudes.  Open squares: $H$ magnitudes.
Open circles: $K$ magnitudes.  Solid line with a large open circle:
our model light curve for the $1.15 ~M_\sun$ WD with $X=0.55$,
$X_{\rm CNO}= 0.10$, $X_{\rm Ne}= 0.03$, and $Z=0.02$ (see Paper I).
The large open circle at the right-lower edge of the line indicates
the end of optically thick wind phase.
  \label{v1500_cyg_yvijhk_color}}
\end{figure}


\begin{figure}
\epsscale{0.8}
\plotone{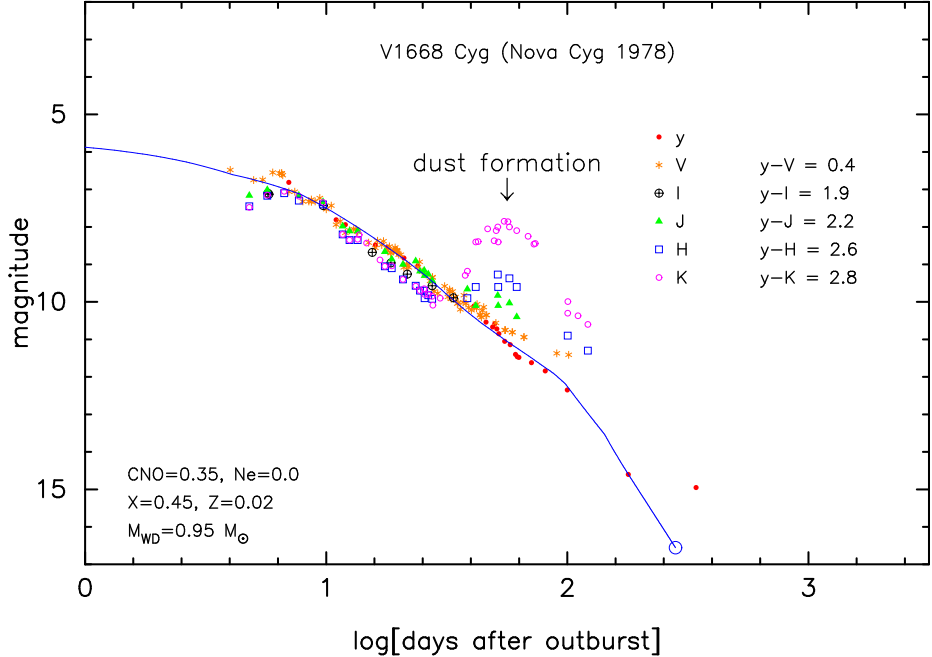}
\caption{
Same as Fig.\ref{v1500_cyg_yvijhk_color} but for the
classical nova V1668 Cyg.  Each light curve is overlapped
with the $y$ light curve between day 6 and 30.
Optically thin dust was formed aound day 35--50. 
Solid line with a large open circle: our model light curve
for the $0.95 ~M_\sun$ WD with $X=0.45$, $X_{\rm CNO}= 0.35$, 
and $Z=0.02$ (see Paper I). 
  \label{v1668_cyg_yvijhk_color}}
\end{figure}


\begin{figure}
\epsscale{0.8}
\plotone{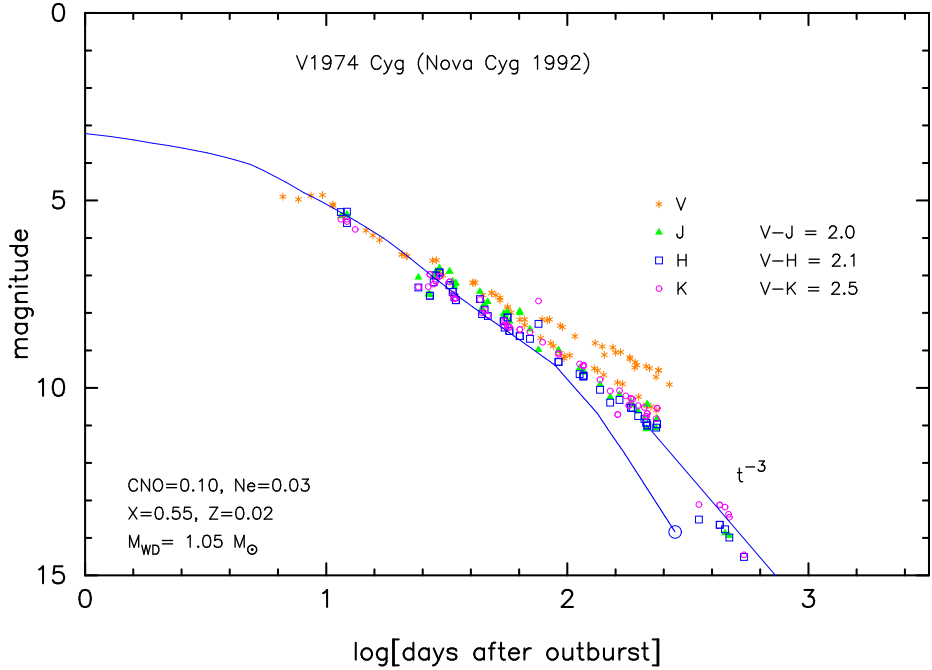}
\caption{
Same as Fig.\ref{v1500_cyg_yvijhk_color} but for the
classical nova V1974 Cyg.  No $y$ and $I$ magnitudes
are available for this nova, so that we show only three
colors of $V-J$, $V-H$, and $V-K$.  Each light curve is
overlapped with the $V$ light curve during day 10--30.
Solid line with a large open circle: our model light curve
for the $1.05 ~M_\sun$ WD with $X=0.55$, $X_{\rm CNO}= 0.10$, 
$X_{\rm Ne}= 0.03$, and $Z=0.02$ (see Paper I).  Solid line
with $t^{-3}$: a decline law of $F_\lambda \propto t^{-3}$
(see eq.[\ref{free-free-stop}]).
  \label{v1974_cyg_vjhk_color}}
\end{figure}


\begin{figure}
\epsscale{0.8}
\plotone{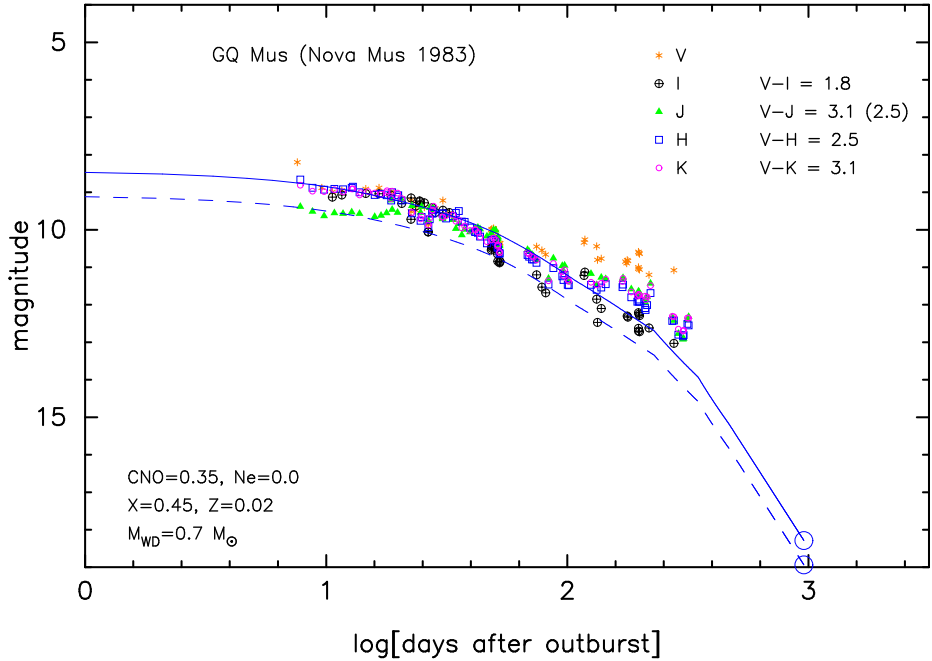}
\caption{
Same as Fig.\ref{v1500_cyg_yvijhk_color} but for the
classical nova GQ Mus.  No $y$ magnitudes are available
for this nova, so that we show four colors of $V-I$,
$V-J$, $V-H$, and $V-K$.  Each light curve is overlapped
with the $V$ light curve during day 8--50 except for the
$J$ magnitudes.  The $J$ magnitudes are fitted only during
day 20--100.  As mentioned in \S \ref{optical_infrared},
the $J$ magnitude brightened up by 0.6 mag after day 15.
If we overlap the $J$ light curve between day 8 and 15,
the color decreases from $V-J =3.1$ to $V-J =2.5$ as shown
in the figure.
Solid line with a large open circle: our model light curv
for the $0.7 ~M_\sun$ WD with $X=0.45$, $X_{\rm CNO}= 0.35$,
and $Z=0.02$ (see \S \ref{light_curve_fitting}).
Dashed line with a large open circle: same as that for the solid
line but vertically shift down by 0.6 mag to match
the $J$ magnitudes during day 8-15.
  \label{gq_mus_vijhk_color}}
\end{figure}


\begin{figure}
\epsscale{0.8}
\plotone{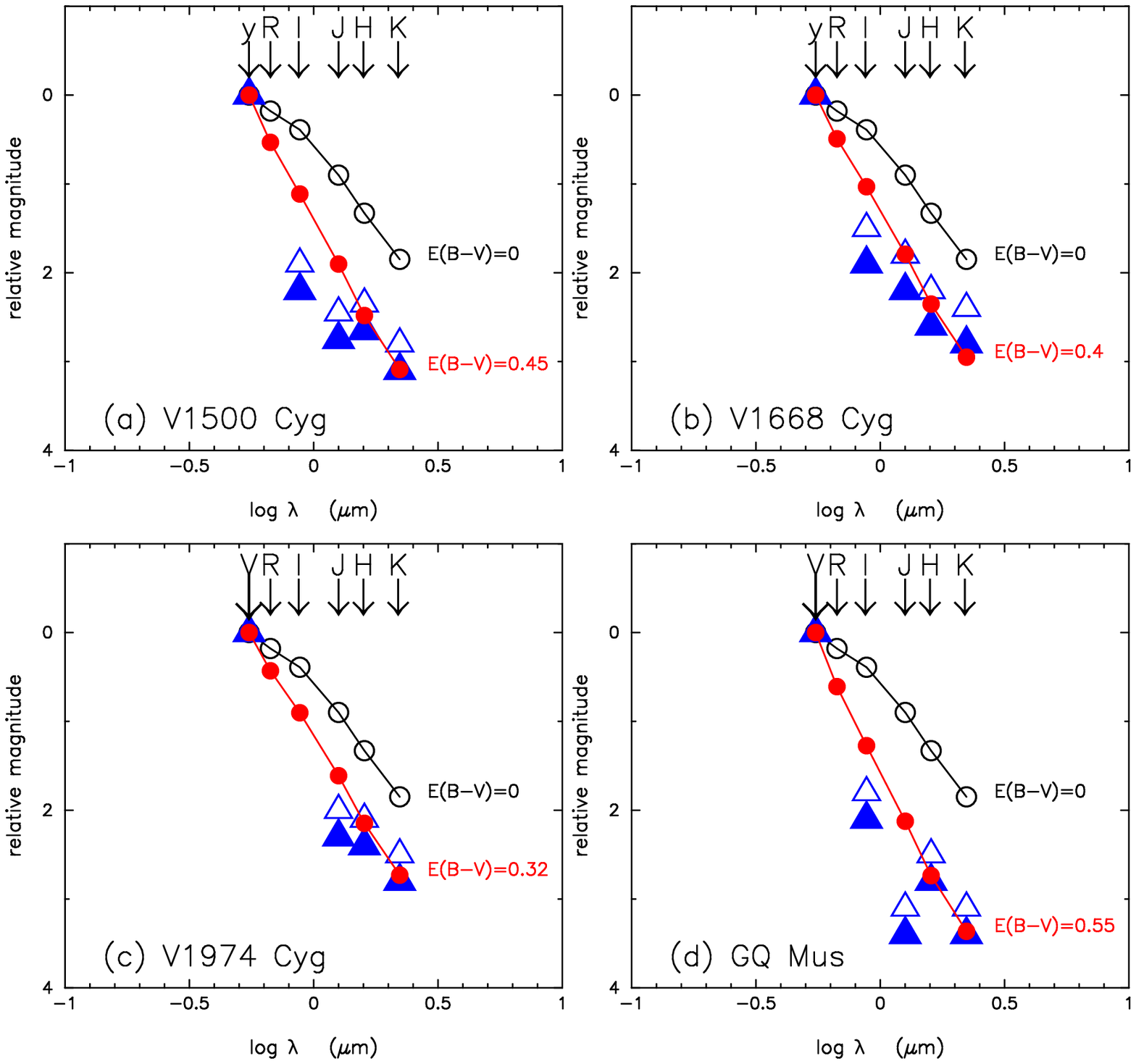}
\caption{
The $V-R$, $V-I$, $V-J$, $V-H$, and $V-K$ ({\it open triangles}) and
the $y-R$, $y-I$, $y-J$, $y-H$, and $y-K$ ({\it filled triangles})
for the four classical novae: (a) V1500 Cyg, (b) V1668 Cyg,
(c) V1974 Cyg, and (d) GQ Mus.  Open and filled circles represent,
respectively, the intrinsic and the reddening corrected colors
as expected from free-free emission (see Table \ref{color_of_free-free}).
See equation (\ref{color_reddening_relation}) for the relation between
free-free color and reddening).  Note that the $H$ and $K$-band is
the least contaminated by emission lines, so that the
$y-H$ and $y-K$ colors are in a reasonable agreement
with the theoretical values.  See text for more details.
  \label{magnitude_calibration3}}
\end{figure}

\section{Color of free-free light curves of novae} 

In this and previous papers we have modeled the time evolution
of novae in the continuum and compared the results with observed
light curves.  Such a study is hampered by the presence of
emission lines, which give rise to sensible distortions in most
photometric bands (see \S 1), especially at some nova phases.
To evaluate the contamination by emission lines, we take
advantage of the fact that the intrinsic colors of free-free
emission are constant with time because of $\lambda F_{\lambda}
\propto \lambda^{-1}$ for free-free emission.  Here, we consider
four classical novae that are well observed in various photometric
bands: V1500 Cyg, V1668 Cyg, V1974 Cyg, and GQ Mus.

Intrinsic $V-R$, $V-I$, $V-J$, $V-H$, and $V-K$ colors from
free-free emission, as calculated from $\lambda F_\lambda \propto
\lambda^{-1}$, are reported in the second column of Table
\ref{color_of_free-free}.  If reddening is known, observed colors
are obtained from
\begin{equation}
m_V - m_\lambda = (M_V - M_\lambda)_0 + c_\lambda E(B-V),
\label{color_reddening_relation}
\end{equation}
where $(M_V - M_\lambda)_0$ is the intrinsic color and
$c_\lambda$ is the reddening coefficient listed in the 
second and third columns of Table \ref{color_of_free-free},
respectively.

In the following subsection, we will compare the observed colors
with those calculated from equation (\ref{color_reddening_relation})
for classical novae V1500 Cyg, V1668 Cyg, V1974 Cyg, and GQ Mus.
Figures \ref{v1500_cyg_yvijhk_color}--\ref{gq_mus_vijhk_color}
demonstrate that the shape of light curves is almost independent
of the wavelength, which is a characteristic feature of free-free
emission (see equation [\ref{free-free-wind}]).  In these figures,
we shift each observed light curve down by $\Delta m_\lambda$ and
overlap it on the observed $y$ (or $V$) light curve.  Then, we
obtain the color of $\Delta m_\lambda = m_y - m_\lambda$
(or $\Delta m_\lambda = m_V - m_\lambda$) from the definition of
$m_\lambda + \Delta m_\lambda = m_y$ (or $m_\lambda + \Delta
m_\lambda = m_V$).  The colors thus obtained are listed in each
figure and plotted in Figure \ref{magnitude_calibration3}.


\begin{deluxetable}{llll}
\tabletypesize{\scriptsize}
\tablecaption{Colors of free-free light curves
\label{color_of_free-free}}
\tablewidth{0pt}
\tablehead{
\colhead{color\tablenotemark{a}} &
\colhead{...} &
\colhead{intrinsic} &
\colhead{coefficient} \\
\colhead{($m_V - m_\lambda$)} &
\colhead{} &
\colhead{($(M_V - M_\lambda)_0$)} &
\colhead{($c_\lambda$)}
}
\startdata
$V-V$ & ... & 0 & 0 \\
$V-R$ & ... & 0.18 & 0.7812  \\
$V-I$ & ... & 0.39 & 1.6058  \\
$V-J$ & ... & 0.90 & 2.2258 \\
$V-H$ & ... & 1.33 & 2.5575 \\
$V-K$ & ... & 1.85 & 2.7528
\enddata
\tablenotetext{a}{color is calculated from
$(m_V-m_\lambda) = (M_V-M_\lambda)_0 + c_\lambda E(B-V)$,
where $(M_V-M_\lambda)_0$ is the intrinsic color of the
free-free spectrum and $c_\lambda$ is the reddening
coefficient}
\end{deluxetable}

\subsection{V1500 Cyg}
For V1500 Cyg, we use the $y$ photometry in \citet{loc76} and the
$V$ data in \citet{tem79}, $I$ data in \citet{the76}.
The $J$, $H$, and
$K$ observations were taken from \citet{enn77}, \citet{kaw76},
and \citet{gal76}.  These light curves are plotted in 
Figure \ref{v1500_cyg_yvijhk_color} with each light curve being
overlapped on the $y$ magnitude.  The spectrum became that for
free-free emission about 4--5 days after the outburst while it
was that for blackbody during the first 3 days \citep{gal76}.  So,
we shift each observed light curve down by $\Delta m_\lambda$
and overlap it on the observed $y$ light curve between day 6 and 30.
Then, we obtain the color of $\Delta m_\lambda = m_y - m_\lambda$
as listed in Figure \ref{v1500_cyg_yvijhk_color}.

The intrinsic color indices, $(M_V - M_\lambda)_0$, for $V-R$, 
$V-I$, $V-J$, $V-H$, and $V-K$ of free-free emission (Table
\ref{color_of_free-free}) are plotted in Figure 
\ref{magnitude_calibration3}a (open circles) together with 
$E(B-V)= 0.45$ (filled circles) calculated from equation 
(\ref{color_reddening_relation}).

The $V-I$, $V-J$, $V-H$, and $V-K$ color indices 
obtained from Figure \ref{v1500_cyg_yvijhk_color}
are plotted in Figure \ref{magnitude_calibration3}a (open triangles).
If strong emission lines are present in the spectrum,
observed colors deviate from those calculated from equation
(\ref{color_reddening_relation}).
In fact, the $V$ band is not emission-line free, being contaminated
by strong emission lines even between day 6 and 30.  The amount
of contamination can be estimated from $y-V= 0.3$ because the $y$
band is almost emission-line free.  This value tells us
that strong emission lines contribute about 30\% of the
energy flux to the $V$ band.  The $y-I$, $y-J$, $y-H$, and $y-K$
color indices are also plotted in Figure 
\ref{magnitude_calibration3}a (filled triangles).

A somewhat larger contamination is present in the $I$ and $J$
bands while the $H$ and $K$ bands are not so heavily contaminated
by emission lines.  For V1500 Cyg, the $y-H$ and $y-K$ colors are
in a reasonable agreement with equation
(\ref{color_reddening_relation}) with $E(B-V)= 0.45$.

\subsection{V1668 Cyg}
Figure \ref{v1668_cyg_yvijhk_color} shows $y$, $V$, $I$, $J$, $H$,
and $K$ light curves of V1668 Cyg.  Here the $y$ magnitude
observations are taken from \citet{gal80}, the $V$ data are
from \citet{mal79} and \citet{due80},
the $I$ magnitudes from \citet{der78},
and the $J$, $H$, and $K$ magnitudes from \citet{geh80}.
Each light curve is shifted down to overlap with the $y$
light curve between day 6 and 30 because the spectrum was no longer
that for free-free emission after day 35 due to dust formation.
The colors are summarized in this figure and are also plotted
in Figure \ref{magnitude_calibration3}b.
The $I$ band is strongly, the $J$ band is slightly, while the
$H$ and $K$ bands are not heavily contaminated by emission lines.
For V1668 Cyg, the $y-H$ and $y-K$ color are in reasonable
agreement with equation (\ref{color_reddening_relation})
together with $E(B-V)= 0.40$ \citep{sti81}.
Note also that the $V$ band is contaminated by emission lines
as suggested from $y-V= 0.4$.

\subsection{V1974 Cyg}
For V1974 Cyg, no $y$ and $I$ photometry is available.  Figure
\ref{magnitude_calibration3}c shows the $V-J$, $V-H$, and $V-K$
colors obtained from Figure \ref{v1974_cyg_vjhk_color},
where $V$ magnitudes are taken from \citet{cho93} and the $J$,
$H$, and $K$ magnitudes are from \citet{woo97}.
Open triangles denote the observed colors, which
follow quite well the calculated ones with $E(B-V)= 0.32$ 
\citep{cho93}.  However, the $V$ band is contaminated
by emission lines even at early phases as already shown in
V1500 Cyg and V1668 Cyg.  If we subtract this
contribution of $\Delta V \sim 0.3$ \citep[estimated from
the spectrum in Fig. 2 of][]{bar93}, i.e., $y-V= 0.3$,
we obtain the corrected colors (filled triangles) as shown
in Figure \ref{magnitude_calibration3}c.  The $J$ band is heavily
contaminated by \ion{O}{1}, Paschen $\beta$, and $\gamma$,
while the $H$ and $K$ bands are not so heavily contaminated
\citep{woo97}.  This explains why the $y-H$ and $y-K$
colors, corrected by $\Delta V = y-V= 0.3$, are in good
agreement with the expected values from free-free emission.

\subsection{GQ Mus}
In Figure \ref{magnitude_calibration3}d we plot (as open triangles)
the observed $V-I$, $V-J$, $V-H$, and $V-K$ colors from Figures
\ref{gq_mus_vijhk_color}, and compare them with
the intrinsic colors from free-free emission
calculated from equation (\ref{color_reddening_relation}) for
$E(B-V)= 0.55$ (filled circles).  The observed colors are
in reasonably good agreement with the calculated ones.

However, the $V$ band is not emission-line free, being contaminated
by strong emission lines even in the early phase \citep[e.g.,
$\sim 30$\% from Fig. 2 of][]{whi84}.  This contributes $\Delta V
= y-V \sim 0.3$ to the colors, so we subtract this from the original
value of $V$, and obtain the corrected colors of $y-I$,
$y-J$, $y-H$, and $y-K$ (filled triangles)
as shown in Figure \ref{magnitude_calibration3}d.  The $I$ band
is strongly contaminated by \ion{O}{1} line and the $J$ band is
also very strongly contributed by Paschen $\beta$ line 
\citep[e.g., Fig. 3 of][]{whi84}.  Consequently, both the $y-I$
and $y-J$ colors (corrected by $\Delta V= y-V=0.3$) 
are 0.8 and 1.3 mag below the calculated ones, respectively.
On the other hand, the $H$ and $K$ bands are not heavily contaminated
by strong emission lines \citep[e.g., Fig. 3 of][]{whi84}.  This
explains why both the $y-H$ and $y-K$ colors corrected by
$\Delta V= y-V=0.3$ are in good agreement with equation
(\ref{color_reddening_relation}) of $E(B-V)= 0.55$.

\end{document}